\documentclass[useAMS,usenatbib]{mn2e}
\usepackage{graphicx}
\usepackage{amssymb}
\usepackage{subfigure}
\usepackage{float}
\usepackage{natbib}
\usepackage{longtable}
\usepackage{xtab}

\title[]{Long-term multiwavelength observations of 1ES\,1218$+$304: physical implications of the flux and spectral variability}
\author[B. Kapanadze et al.]{B. Kapanadze$^{1,2,3}$ \thanks{E-mail:
bidzina\_kapanadze@iliauni.edu.ge}, A. Gurchumelia$^{2,4}$, M. Aller$^{5}$\\
$^{1}$ Ilia State University, Colokashvili Av. 3/5, Tbilisi, Georgia, 0162\\
$^{2}$E. Kharadze National Astrophysical Observatory, Mt. Kanobili, Abastumani, Georgia, 0803 \\
$^{3}$INAF, Osservatorio Astronomico di Brera, Via E. Bianchi 46,
23807 Merate, Italy\\
 $^{4}$I. Javakhishvili State
University, Chavchavadze Av. 3, Tbilisi 0128, Republic of Georgia\\
 $^{5}$Astronomy Department, University of Michigan, Ann
Arbor, MI 48109-1107, USA }
\begin{document}
\date{Accepted . Received ; }
\pagerange{\pageref{firstpage}--\pageref{lastpage}} \pubyear{}
\maketitle

\label{firstpage}
\begin{abstract}
This paper presents the results of a detailed  timing and spectral analysis of the TeV-detected blazar 1ES\,1218$+$304, focused on the observations performed with the different instruments onboard the \emph{Neil Gehrels Swift Observatory} in the period 2005--2024. The source showed various strengths of X-ray flaring activity and 0.3-10\,keV states differing by a factor up to 20 in brightness,  exceeding a level  of 2.7$\times$10$^{-10}$erg\,cm$^{-2}$s$^{-1}$  and representing the 3rd brightest blazar during the strongest  flare. We detected tens of intraday variability instances, the majority of which occurred on sub-hour timescales and were consistent with the shock-in-jet scenario. The spectral properties were strongly and fastly variable, characterized by a frequent occurrence of very hard photon indices in the 0.3--10\,keV and \emph{Fermi} 0.3--300\,GeV bands. The source exhibited very fast transitions  of logparabolic-to-powerlaw spectra or conversely, possibly caused by changes of magnetic field properties over small spatial scales or by turbulence-driven relativistic magnetic reconnection. We detected various spectral features, which demonstrate the importance of the first-order Fermi mechanism operating by the magnetic field of changing confinement efficiencies and by the electron populations with different initial energy distributions, stochastic acceleration and cooling processes.  In some periods, the source showed a softening at higher GeV-band energies, possibly due to the inverse-Compton upscatter of X-ray photons in the Klein-Nishina regime reflected in the positive correlation between X-ray and high-energy emissions.
\end{abstract}


\section{Introduction}
BL Lacertae objects (BL Lac) form  a blazar subclass which, in turn, represent active galactic nuclei (AGNs) notable for strong flux variability on timescales from a few minutes to years across all wavelengths, high polarization, and superluminal speeds of some radio-components \citep{f14}. These properties are commonly explained to stem from  a relativistic jet pointed to the observer (see, e.g., \citealt{pad17}). BL Lacs also show a very extended spectral energy distribution (SED), which is stretched over 17--19 orders in frequency  and consists of the two smooth, broad distinct components in the $\nu F_{\rm \nu}$ representation. The first "hump" extends from the radio frequencies to X-rays and  is firmly explained  to be synchrotron emission from (ultra)relativistic electrons (and, possibly, positrons) populating the jet emission zone (see, e.g., \citealt{pad17}). Depending on the position of the corresponding peak, BL Lacs are sub-classified as low, intermediate and high-energy-peaked BLLs (LBL, IBL and HBLs, respectively; \citealt{p95,b01}). The latter group exhibits the synchrotron SED peak at the UV--X-ray frequencies and form a vast majority (more than 80\,per cent) of the extragalactic TeV-detected sources\footnote{see http://tevcat.uchicago.edu}, hinting at the presence of the highest-energy particles and the most violent acceleration processes in their jets \citep{ah09}. However, the origin of the higher-energy  "hump" is still disputable, and  there are two "competing" models developed for this purpose. The first scenario incorporates an inverse Compton (IC) scattering of low-energy photons by the  "parent" electron population (synchrotron self-Compton model, SSC; \citealt{m85}), or the "seed" photons may have an external origin (external Compton model; \citealt{d92}). Alternatively, hadronic models explain the $\gamma$-ray emission by be produced relativistic protons via the synchrotron mechanism in a highly-magnetized jet medium (the proton-synchrotron model and its modifications; \citealt{m93,c15}), as well as by synchrotron emission from a secondary electron population produced by the interaction between the high-energy protons and ambient photons (see, e.g., \citealt{z17}).  For the specific HBL source, the suitable model can be selected  and conclusions about the unstable processes can be drawn through the multiwavelength (MWL) flux variability and correlation studies, as well as by investigating the X-ray and  $\gamma$-ray spectral properties. The latter study can also resolve  the problems associated with the jet particle content, their acceleration up to ultrarelativistic energies, and unstable mechanisms  underlying the observed MWL variability. By comparing the  results obtained for different HBLs, we can draw statistical conclusions about the importance of the aforementioned emission scenarios and possible unstable processes operating in the HBL jets.

We have performed such a study for the relatively poorly investigated (compared to the nearby X-ray and $\gamma$-ray bright blazars),   TeV-detected HBL  source 1ES\,1218$+$304  focused on the long-term observations performed with the X-Ray Telescope (XRT,  \citealt{bur05}) onboard the satellite \emph{Swift} \citep{g04}  during 2005--2024.  In order to check the  inter-band cross-correlation and lognormality of the MWL flux variability, we included the data obtained with the (1) Ultraviolet-Optical Telescope (UVOT; \citealt{r05}) and Burst Alert Telescope (BAT; \citealt{ba05}) based on the same satellite; (2) Monitor of All Sky X-ray Image (MAXI; \citealt{m09}); (3)  Large Area Telescope (LAT) onboard the satellite \emph{Fermi} \citep{at09}; (4) ground-based telescopes. 

The paper is organized as follows: Sections 2 presents the highlight of the past studies of the target. Section\,3 describes the data processing and analysis procedures. In Section\,4, we provide the results of the X-ray timing analysis and  spectroscopy, along with the temporal study of the data sets available in other spectral bands. Based on these results, the corresponding discussion, conclusions and  physical interpretation is presented in Section\,5. Finally, the summary of our study is given in Section\,6.

\section{Past studies}

The source was originally detected at 408 MHz within the Bologna Northern Cross Telescope Survey (B2; \citealt{c70}). The X-ray counterpart was identified  within the \emph{Ariel All Sky Survey} \citep{c78}). The first identification as a possible BL Lac source was made by \cite{w79}, and it was denoted as 1ES\,1218$+$304 in Eintein Slew Survey (1ES, \citealt{e92}). The redshift z=0.182 was determined by \citep{b98} via the  optical spectrum obtained at the Calar Alto 3.5\,m telescope. The optical imaging with the Hubble Space Telescope showed the presence of elliptical host galaxy \citep{f99}. 

The source showed both curved (logparabolic) and power-law  spectra, as well as a flux variability on long-term and intraday timescales during the X-ray observations performed with \emph{BeppoSAX} (1999 July; \citealt{g05}), \emph{XMM-Newton} (2001\,June; \citealt{m08,d22}),  \emph{Swift}-XRT  (2006\,March--May; \citealt{t07,m08}), \emph{Suzaku} (2006 May; \citealt{s08}),  \emph{NuSTAR} (2015 November; \citealt{pan18,c18}) and \emph{AstroSat} (2019 January; \citealt{d23}).  X-ray flaring states with hard 0.3--10\,keV spectra during 2019 January, 2023 June--July and 2024 November was alerted by \cite{fal19} and \cite{k23a,k24a}. Although the results from the Swift, MAXI and LAT observations from the period of 2008 September--2018 August were presented by \cite{s19}, the study was restricted to the \emph{Swift} exposures longer than 1\,ks, power-law spectral model by using only the neutral hydrogen column density and 0.1--300\,GeV energy range for the LAT  observations (while the  range of 0.3--300\,GeV is generally recommended for the HBL sources; see, e.g., \citealt{ab11}); no cuts on the detection significances were made in the case of the BAT and  MAXI observations. Moreover, (i) \cite{sah20} included the data from the MWL carried out during 2008--2020, but the \emph{Swift}-XRT data (only those obtained from the photon-counting observations) were not processed by means of the commonly-adopted technique; the same source extraction and pileup estimation radii used for all selected observation; the results were presented for the 0.3--7\,keV band (instead the entire 0.3--10\,keV energy range); no information about the XRT spectral analysis and flux extraction procedure was provided; the  range of 0.1--300\,GeV for the LAT data and the source detection significance of 2$\sigma$ adopted; no common practice of the de-reddening of the UVOT-band fluxes adopted; (ii) \cite{d23} reported the results from the near-simultaneous data obtained during the period from 2018\,January--2021\,May. However, the 10-sec source radius was adopted for all XRT observations, while this object generally requires significantly larger radii (see below); only the power-law spectral model and the neutral hydrogen column density were adopted. 

\begin{table*} \centering \small
  \begin{minipage}{170mm}
\caption{\label{xrttable} Summary of the XRT observations (extract):  column (1) -- observation ID (the three leading zeroes are omitted throughout the paper); (2) -- observation start and end times (in UTC); (3) -- exposure (in seconds); (4) -- observation mode; (5) -- Modified
Julian Date; (6) -- mean value of the 0.3--10\,keV count rate (CR, in cts\,s$^{-1}$) and the corresponding error shown within the parenthesis; (7)--(9): reduced  Chi-squared with the corresponding degrees-of-freedom (DOF), time bin (in seconds) and  existence of the flux variability (NV: non-variable), respectively.} \vspace{-0.2cm} \centering
  \begin{tabular}{cccccccccc}
         \hline
 ObsID & Obs. Start--End (UTC) & Exp. (s)&Mode &MJD (d) & CR(cts\,s$^{-1}$) & $\chi^2_{\rm r}$/DOF.& Bin\,(s)&  Var.  \\
(1) & (2) & (3) & (4) & (5) & (6) & (7) & (8)&9 \\
\hline
35016002&	2005-10-30 21:17:01--10-30 23:14:56&	53673.882	&2032&	PC	&1.79(0.04)	&1.13/16	&120&	NV \\
35016001&	2005-10-31 19:48:02--10-31 23:24:59	&53674.825&	3772&	PC	&1.57(0.03)	&0.95/31&	120	&NV \\
30376001&	2006-03-08 03:10:00--03-08 05:16:00	&53802.132&	3105&	WT	&0.87(0.02)	&0.99/26	&120	&NV \\
\hline
\end{tabular}
\end{minipage}
\vspace{-0.2cm}
\end{table*}

After the original TeV-detection with MAGIC  in 2005\,January \cite{al06}, the VERITAS observations showed  a factor of $\sim$5 higher flux (compared to that recorded about 2\,yr earlier) and a very hard spectrum with $\Gamma$$\sim$1.5 in 2009 January \citep{ac10}. A similar spectrum was also reported from the LAT 0.2--300\,GeV observations \citep{ab09}.  The source was found in a VHE flaring state with MAGIC and  VERITAS during 2018\,December\,31--2019\,January\,2 \citep{myr19,muk19}. Simultaneously, the optical $R$-band and infrared brightnesses attained the  15-yr highest level \citep{myr19,car19}. Later, the source showed an even higher $R$-band state in 2020 April \citep{j20}. \cite{c18,c20} included our target in the list of TeV-peaked candidate BL Lacs with the higher-energy SED peak possibly situated beyond 2\,TeV.

As for other studies, the optical monitoring with the ATOM telescope and  Tuorla blazar  program\footnote{https://users.utu.fi/kani/1m/} during 2002--2012 showed a bluer-when-brighter behavior of the source, overall variability by 1.0--1.2 magnitude and   power-law PSD (power spectral density; \citealt{w15,n18}). The OVRO blazar monitoring program\footnote{https://sites.astro.caltech.edu/ovroblazars/} carried out at 15\,GHz proved 1ES\,1218$+$304 to be one of the faintest radio sources among the Northern TeV-blazars and very weakly variable \citep{l16}. 
  
\section{DATA SETS, REDUCTION and ANALYSIS }

\subsection{X-ray observations}
The XRT data from the XRT observations of 1ES\,1218$+$304 were retrieved from   NASA’s Archive of Data on Energetic Phenomena (HEASARC\footnote{http://heasarc.gsfc.nasa.gov/docs/archive.html}).  The source was observed 180 times with  between 2005\,October\,30 and 2024\,November\,27.  The raw event files  were processed and reduced by using the script \texttt{XRTPIPELINE} incorporated by \texttt{XRTDAS} package (to be a part of \texttt{HEASOFT}). For these purposes,  we adopted the standard grade, region, time, energy and phase filtering criteria, as well the latest calibration files. Moreover, the bad pixels (dead, hot or flickering) and those below the event threshold for the Earth limb screening were removed.  

The majority of these snapshots (about 60\%) were performed in the Windowed Timing (WT) mode, and the remaining ones in the Photon Counting (PC) mode (see Table\ref{xrttable}). The source and background extraction regions were generated by using the tasks incorporated in the package \texttt{XSELECT}. Generally, the core of the target's point spread function (PSF) is affected  by photon pile-up for fluxes $\gtrsim$0.5 cts\,s$^{-1}$ in the PC mode. Therefore,  the source events were extracted from an annular region with an inner radius of 2--10 pixels and an outer radius of 15--55 pixels, depending on the target's brightness and exposure duration.  The size of a piled-up area was estimated by fitting the  PSF wings with the analytical model developed by \cite{m05}. The target's light curve was then corrected for the resultant loss of telescope effective area, bad/hot pixels, pile-up, and vignetting by means of the task \texttt{XRTLCCORR}. The background counts were extracted from the concentric area, centered on the source, with inner and outer radii of 80 and 120 pixels, respectively. Using the \texttt{XRONOS} task \texttt{LCURVE}, the background-subtracted 0.3-10\,keV light curves was constructed by adopting various time bins (from 30 to  390 seconds, in attempt to detect a fast variability).  In order to extract the 0.3--10\,keV spectrum,  we generated the corresponding ancillary response (ARF) file by means of the task \texttt{XRTMKARF} with corrections for the PSF losses, different extraction regions, CCD defects and vignetting.

As for the WT-mode observations, the light curves and spectra were generated by using the circular areas with radii of 10--40\,pixels.  A significant fraction of the XRT observations were performed during two or more different \emph{Swift} orbits, and the corresponding WT images generally do not have the common CCD pixel with the highest count. Since such a situation poses a risk of obtaining incorrect timing and spectral information, we split the original event file into the parts corresponding to a single \emph{Swift} orbit and then performed the aforementioned analysis separately.

\begin{table*}
\tabcolsep 1.5pt   \small
  \begin{minipage}{180mm}
  \caption{\label{uvot} The results of the UVOT observations (extract). Columns (2), (4), (6), (8), (10) and (12) provide the corrected magnitudes in each band, and the corresponding flux values (in   mJy) are given in the next columns, respectively.} \vspace{-0.2cm} \hspace{-1.0cm}
  \begin{tabular}{ccccccccccccccc}
       \hline
ObsID & \multicolumn{2}{c}{\emph{V}} & \multicolumn{2}{c}{\emph{B}} & \multicolumn{2}{c}{\emph{U}} & \multicolumn{2}{c}{\emph{UVW1}} & \multicolumn{2}{c}{\emph{UVM2}}& \multicolumn{2}{c}{\emph{UVW2}}\\
\hline
(1) & (2) & (3) & (4) & (5) & (6) & (7) & (8) & (9) & (10) & (11) & (12) & (13)\\
\hline
35016002&	16.15(0.07)&	1.04(0.07)&	16.29(0.05)	&1.18(0.07)&	15.30(0.04)	&1.09(0.04)&	15.12(0.04)&	0.80(0.03)	&15.00(0.05)&	0.77(0.03)&	15.08(0.04)	&0.69(0.03)\\
35016001&	15.83(0.07)&	1.48(0.10)&	16.37(0.05)&	1.09(0.07)&	15.65(0.04)&	0.78(0.03)&	15.43(0.04)&	0.60(0.02)	&15.37(0.05)&	0.54(0.02)&	15.42(0.04)&	0.50(0.02)\\
30376001&	16.23(0.06)&	0.95(0.06)&	16.63(0.06)&	0.84(0.04)&	15.61(0.04)	&0.81(0.03)&	-&	-&	-	&-&	-	&-\\
 \hline \end{tabular} \end{minipage} \end{table*}

We retrieved the daily-binned BAT-band fluxes from the website of the program Hard X-ray Transient Monitor \footnote{See http://swift.gsfc.nasa.gov/results/transients/weak/QSOB1218p304} \citep{kr13}. Note that the BAT is a coded-mask device and, therefore, only the detections with a 5$\sigma$ significance are generally used for the variability studies (see, e.g., \citealt{k24b}). However, 1ES\,1218$+$304 was rarely detectable with this significance, and we re-binned the orbit-resolved BAT data  by using the time intervals from 2 days to several weeks via the \texttt{HEASOFT} task \texttt{REBINGAUSSLC}. 

MAXI also is a coded-mask device, and we filtered the background-subtracted, 1-day-binned 2--20\,keV count rates (retrieved from the mission's website\footnote{http://maxi.riken.jp}) using the 5$\sigma$ detection threshold for the flux variability study. In attempt to increase the number of detections with this significance, we used the online tool \texttt{MAXI ON-DEMAND PROCESS} (available on the same website) and derived the 2--20\,keV light curve with the integration of 2 days and longer. 

\subsection{UV--optical data}

During each \emph{Swift} visit to our target, the UVOT mostly performed observations in the optical (\emph{V, B, U}) and  UV (\emph{UVW1, UVM2, UVW2} bands. Using the sky-corrected images (also retrieved from HEASARC), photometric measurements were carried out  via the  \texttt{HEASOFT} tool \texttt{UVOTSOURCE}. We created a source extraction region of the 4--5 arcsec and 8--10 arcsec radii in the optical and UV bands, respectively. The  background regions of the same size were created sufficiently away from the source.  The optical--UV magnitudes were then corrected for Galactic absorption applying $E(B-V)$=0.028\,mag (derived according to  \citealt{g09}) and  the $A_\lambda/E(B-V)$ values, which were calculated using the interstellar extinction curves provided by \cite{fi07}. Consequently, the quantity $A_\lambda$ is equal to 0.09, 0.11, 0.14, 0.15, 0,22 and 0.20 magnitudes in the bands \emph{V--UVW2}, respectively. The corrected magnitudes and fluxes are provided in Table\,\ref{uvot}.  

 We also used the publicly-available optical \emph{R}-band  data  obtained (i) within the Tuorla program and published by \cite{n18}\footnote{http://cdsarc.u-strasbg.fr/viz-bin/qcat?J/A+A/620/A185}; (ii) with the 0.76-m Katzman Automatic Imaging Telescope (KAIT; \citealt{li03}) and available at KAIT Fermi AGN Light-Curve Reservoir\footnote{http://herculesii.astro.berkeley.edu/kait/agn/}; (iii) with the 2.3-m Bock and 1.54-m Kuiper telescopes of Steward
Observatory (see Smith et al. 2009 for details). The results are available on the corresponding website\footnote{https://james.as.arizona.edu/\%7Epsmith/Fermi/)}, which also contains the \emph{V}-band  data. The latter were also obtained in the framework of the Catalina Real-time Transient Survey\footnote{http://nesssi.cacr.caltech.edu/catalina/Blazars/Blazar.html}. 

The host contribution was estimated by adopting the optical colour-indices from \cite{f95} and \emph{R}-band value $m_{\rm R}$=17.21\,mag (corresponding to 0.40\,mJy; \citealt{n18}). In the \emph{V, B} and \emph{U} bands, this quantity amounts to 0.23\,mJy, 0.06\,mJy and 0.01\,mJy, respectively. As for the UV fluxes in the \emph{UVW1--UVW2} bands, they are not significantly affected by the host emission.

\subsection{Gamma-ray data}

The Fermi-LAT data were extracted from the LAT Data Server\footnote{https://fermi.gsfc.nasa.gov/cgi-bin/ssc/LAT/LATDataQuery.cgi} and analysed by using the \texttt{Fermitools} package (version 2.2.0\footnote{http://fermi.gsfc.nasa.gov/ssc/}). During this process, we adopted the instrument response function \texttt{P7SOURCE\_V7} and extracted the 0.3--300\,GeV fluxes from the LAT observations: as 1ES\,1218$+$304 is a relatively hard $\gamma$-ray source, its 100--300\,MeV emission is heavily overshadowed by those of the background point sources. We adopted the data cuts as follows: (i) only the events of the diffuse class (characterized by the highest probability to be associated to the source) from a 10\,deg region of interest (ROI)  centered on our target were used; (ii) \texttt{evclass}=128 and \texttt{evtype}=3: the spacecraft was situated outside the South Atlantic Anomaly and the target -- within the LAT FOV; (iii) contamination from the Earth-albedo $\gamma$-rays was reduced by cutting the zenith angles higher than 90\,deg; (iv)  spacecraft's rocking angles lower than 52\,deg were adopted (time intervals when the Earth was outside the LAT FOV). We created a background model via the tool \texttt{make4FGLxml.py}\footnote{https://fermi.gsfc.nasa.gov/ssc/data/analysis/user/make4FGLxml.py}. This model  contains all point-like $\gamma$-ray sources included in the 4th \emph{Fermi}-LAT catalogue (4FGL, \citealt{ab20}) within a radius of 20\,deg centered on the target, as well as the Galactic and extragalactic diffuse  components. Similar to this catalogue,  a simple power-law model for the spectral modeling of 1ES\,1218$+$304  and extraction of the photon flux was adopted. During this process, the spectral parameters of sources within the ROI were left free during the minimization process (except for those showing a detection significance lower than 3$\sigma$ for the given integration time), while those beyond this range were fixed to the 4FGL values. When the source was not detectable  with at least 3$\sigma$ significance and/or \texttt{GTLIKE} yielded a low value for the model-predicted photons $N_{\rm pred}\lesssim$8, we calculated the upper limit to the photon flux using the Python-based tool \texttt{UpperLimits} \footnote{fermi.gsfc.nasa.gov/ssc/data/analysis/scitools/upper\_limits.html}. Note that a detection even with the significance higher than  3$\sigma$ is not robust when $N_{\rm pred}$ is below this threshold: even a slight change in the bin width yields  significantly different values of the photon flux and spectral parameters (see, e.g., \citealt{k20,k24b}).

We also checked the nightly-binned, quick-look VHE data obtained the FACT telescope \citep{a13} and publicly-available on the instrument's website \footnote{See http://www.fact-project.org/monitoring}. However, these  data did not show the target's detection with a minimum significance of 3$\sigma$, and we have not used them in our study.

\begin{table*} \tabcolsep 3.5pt \small  \begin{minipage}{185mm}
\vspace{-0.5cm}
\hspace{-1cm}
\caption{\label{xrtper} Summary of the XRT observations in the specific periods. The top part: the MJD range (column 2);   the maximum and mean values of the 1-day-binned 0.3--10\,keV count rate (cts\,s$^{-1}$), along with the fractional variability amplitude (columns 3--5); the maximum, mean and minimum values of the unabsorbed 0.3--10\,keV (in 10$^{-11}$erg\,cm$^{-2}$s$^{-1}$), columns 6--8) and photon index at 1\,keV (columns 9--11). The bottom part:   percentage of the power-law spectra (column 2); the maximum, mean and minimum values of  the 0.3--10\,keV photon index (columns 3--5), curvature parameter (columns 6--8) and the position of the synchrotron SED peak (columns 9-11).}
  \vspace{-0.2cm}
  \centering
 \hspace{-0.7cm}
  \begin{tabular}{ccccccccccccccccccc}
    \hline
   &   &\multicolumn{3}{c}{CR(XRT)}  &  \multicolumn{3}{c}{$F_{\rm 0.3--10\,keV}$}    &   \multicolumn{3}{c}{$a$}  \\
    \hline
Per. &MJDs & Max & Mean & $F_{\rm var}(\%)$ &Max & Min  & Mean & Max   & Min   & Mean     \\
(1) & (2) & (3) & (4)& (5)& (6) & (7) & (8)& (9)& (10)& (11)  \\
 \hline
1  &55970--56100 &4.44(0.07)&2.00(0.02)&46.6(1.7)&14.79(0.37)&1.97(0.15)&8.79(0.08)&2.13(0.05)&1.80(0.05)&2.02(0.02)  \\
2 &56270--56820 &2.55(0.06)&1.66(0.01)&22.6(0.6)&8.17(0.42)&3.40(0.20)&5.27(0.05)&2.15(0.05)&1.60(0.07)&1.92(0.02)  \\
3 &56890--57550 &1.79(0.06)&1.09(0.01)&35.0(0.8)&7.16(0.46)&1.14(0.10)&3.37(0.04)&2.01(0.09)&1.74(0.06)&1.85(0.02)  \\
4 &57715--58255 &3.09(0.09)&1.51(0.01)&47.6(0.8)&11.70(0.55)&1.36(0.09)&4.99(0.05)&2.00(0.05)&1.68(0.08)&1.86(0.03)  \\
5 &58430--59025 &5.71(0.05)&3.37(0.02)&42.3(0.6)&27.69(1.05)&3.69(0.14)&12.36(0.07)&2.46(0.07)&1.53(0.08)&1.96(0.01) \\
6 &59575--60180 &8.00(0.21)&3.35(0.01)&48.6(0.6)&25.79(1.22)&3.48(0.18)&10.78(0.05)&2.01(0.05)&1.59(0.06)&1.82(0.01)  \\
7& 60420--60650&3.86(0.06)&2.85(0.02)&38.7(0.6)&12.00(0.46)&5.14(0.16)&8.59(0.07)& 2.04(0.06)&1.67(0.05)&1.87(0.01)  \\
 \hline
  & &\multicolumn{3}{c}{$\Gamma$}  &   \multicolumn{3}{c}{$b$}   &    \multicolumn{3}{c}{$E_{\rm p}$} \\
    \hline
 Per. & $N_{\rm pow}$(\%)& Max& Min  & Mean & Max & Minn   & Mean & Max  & Min   & Mean \\
(1) & (2) & (3) & (4)& (5)& (6) & (7) & (8)& (9)& (10)& (11)  \\
 \hline
1 &59 &2.17(0.09)&1.88(0.04)&1.98(0.01)&0.37(0.13) &0.22(0.09) &0.28(0.05) &2.85(0.25)&0.51(0.14)&1.15(0.05) \\
2 &46 &2.21(0.05)&1.73(0.05)&1.97(0.01)&0.43(0.15)&0.19(0.11)&0.29(0.04)&4.22(0.36)&0.65(0.12)&1.70(0.05)  \\
3 &69 &2.27(0.08)&1.69(0.07)&1.99(0.02)&1.06(0.37)&0.21(0.12)&0.43(0.06)&4.16(0.36)&0.97(0.27)&2.03(0.08)  \\
4 &67 &2.13(0.06)&1.68(0.05)&1.98(0.02)&0.85(0.44)&0.28(0.12)&0.49(0.08)&2.44(0.24)&1.00(0.13)&1.52(0.09)  \\
5 &49 &2.690.07)&1.60(0.06)&2.00(0.01)&0.73(0.19)&0.20(0.12)&0.37(0.03)&4.32(0.37)&0.21(0.13)&1.46(0.04) \\
6 &42 &2.18(0.05)&1.71(0.08)&1.94(0.01)&0.72(0.14)&0.17(0.10)&0.32(0.02)&4.83(0.42)&0.96(0.12)&2.23(0.03) \\
7&38 &2.15(0.05)&1.87(0.04)&2.02(0.01)&0.61(0.17)&0.18(0.10)&0.35(0.03)&4.64(0.41)&0.87(0.16)&1.81(0.05)  \\
 \hline \end{tabular} \end{minipage} \end{table*}

\section{Results}

\subsection{X-ray and MWL variability on different timescales}
The \emph{Swift} observations of 1ES\,1218$+$304 show a large overall X-ray variability during the 19\,yr period  (see   Table\,\ref{xrttable}). Namely, a wide range of the 0.3--10\,keV count rate  showed an overall change by a factor $\sim$20 with the maximum value CR=8.00(0.21)\,cts\,s$^{-1}$ (corresponding to the unabsorbed flux of about 2.8$\times$10$^{-10}$erg\,cm$^{-2}$s$^{-1}$). In the periods of the relatively intense XRT monitoring, the source exhibited different variability amplitudes on timescales from a few months (strong flares by factors of 4--6) down to several days, and intra-day brightness fluctuations (i.e. those on timescales shorter than 1\,d) were frequently observed, which sometimes occurred within a few hundred seconds (see below). On the contrary, only a moderate variability was observed during a few weeks in some epochs.   Based on the maximum and mean fluxes, as well as on the  $F_{\rm var}$ value, we have discerned seven time intervals ("periods"), the summary of which are presented in Table\,\ref{xrtper} and discussed in detail (along with the target's  timing behaviour in other spectral bands) in Appendix\,A.  This table provides the maximum and mean fluxes (both in 0.3--10\,keV count rates and unabsorbed flux), the maximum-to-minimum flux ratio and the fractional variability amplitude. The latter quantity was calculated as \citep{v03}
\begin{equation}
  F_{\rm var}=\Biggl\lbrace {S^2-\overline{\sigma^2_{\rm err}} \over{F_{\rm mean}}} \Biggr\rbrace ^{1/2}, \\
 \vspace{-0.3cm}
\end{equation}
with  $S^2$, the sample variance; $\overline{\sigma^2_{\rm err}}$, the mean square error, and $F_{\rm mean}$, the mean flux. The XRT-band and MWL light curves in each period are presented in Fig\,\ref{mwlper}. In the latter, the periods are arranged in order from the strongest (Period\,6) to gradually decreasing XRT-band activity. 

Moreover, the source showed 12 flux doubling/halving instances (taking into account the uncertainties; see Table\,\ref{doubl}).  Note that the corresponding timescale $\tau_{\rm d,h}$ was calculated as follows \citep{sa13}
\begin{equation}
 \tau_{\rm d,h}=\Delta t \times ln(2)/ln|(F_{\rm f}/F_{\rm i})|, 
\end{equation}
with $F_{\rm i}$ and $F_{\rm f}$ to be the initial and final flux values, respectively, and $\Delta t$, the corresponding maximum duration. 

   \begin{figure*}
\includegraphics[trim=6.0cm 2.4cm 0cm 0cm, clip=true, scale=0.88]{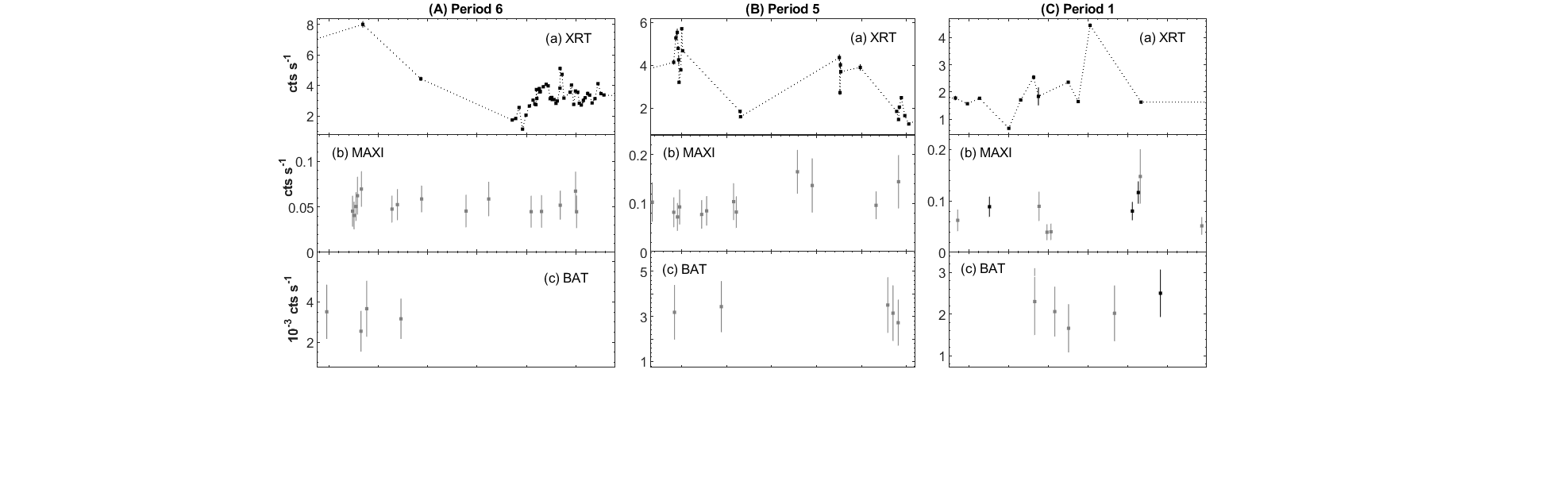}
\includegraphics[trim=6.0cm 1.8cm 0cm 0.1cm, clip=true, scale=0.88]{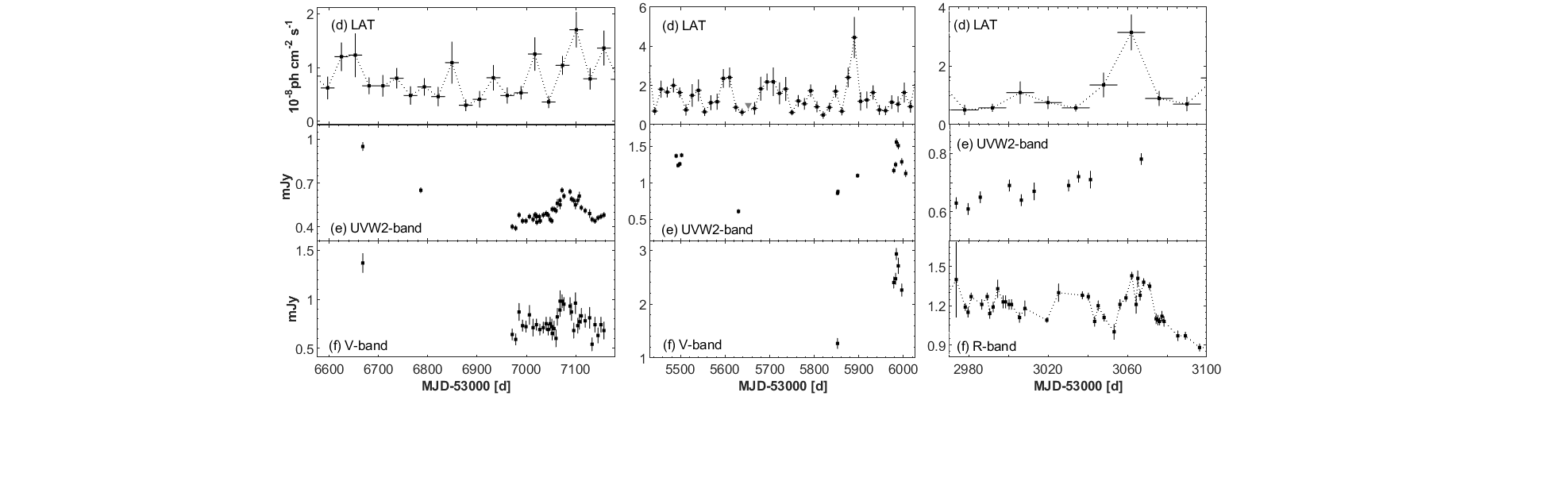}
\vspace{-0.6cm}
 \caption{\label{mwlper}  The MWL variability in different periods listed in Table\,\ref{xrtper} (daily bins: the XRT, MAXI, UVOT, and \textit{R}-band data;  2\,days: BAT;  LAT, two (plots B--E) or four (A and F--G) weeks). The gray and black points in the BAT and MAXI-band plots correspond to the detections with the (3--4)$\sigma$ and 5$\sigma$ significances, respectively. The gray triangles in the LAT-related plots (next page) show the upper limits to the 0.3--300\,GeV photon flux when the source was not robustly detected. }  \vspace{-0.3cm}
 \end{figure*} 

During the XRT observations, the source underwent an  0.3-10\,keV intraday variability (IDV)  at the  3$\sigma$ confidence level 38-times, and the  corresponding summary is provided in Table\,\ref{idvtable}. The latter lists  the total length of the specific observation, $F_{\rm var}(\%)$  and the ranges of different spectral parameters. Fig.\,\ref{idv} presents the IDVs occurring on sub-hour timescales. Namely, the instances characterized by the flux variability within the 1-ks time interval  are given in Figs\,\ref{idv}a--\ref{idv}k and discussed in detail in Appendix\,A.   Other IDVs occurred  during those observations which were distributed over two or more XRT orbits, and the corresponding details are provided in Table\,\ref{idvtable}. These and above discussed IDVs were recorded in different X-ray states with the observation-binned count rate $\overline{CR}$=0.52(0.03)--5.71(0.05)\,cts\,s$^{-1}$.
 
The 1-day-binned MAXI data yielded the maximum number of the target's detections with  5$\sigma$ significance (17 times), while this number dropped significantly along with the wider time bins (1--6 detections in the case of the 2--4 day and 1-week time integrations). Along with these detections, we also plotted those  corresponding to the detection with (3--4)$\sigma$  significances (included in the flux variability study by \citealt{d23}) in Fig.\,\ref{xrtper}, in order to discerning the time intervals of relatively enhanced hard 2--20\,keV activity. Note that these detections are distributed relatively evenly with time during the entire period of the MAXI operations, while the  5$\sigma$-detections on the daily timescales did not occur after 2017 April 27 (MJD\,57871; Period\,4). Afterwards, such detections were recorded twice only in the case of the larger time integrations (during MJD 58488--59494 and 60068--6071; Periods 5 and 6, respectively). Note that  the 5$\sigma$-detections did not exhibit any long-term trends, and the highest 2--20\,keV state was recorded on MJD\,57585 (the time interval between  Periods 3 and 4) when the source was not monitored with XRT and UVOT. 

 \addtocounter {figure}{-1} \begin{figure*} \vspace{-0.2cm}
\includegraphics[trim=5.9cm 2.4cm 0cm 0cm, clip=true, scale=0.88]{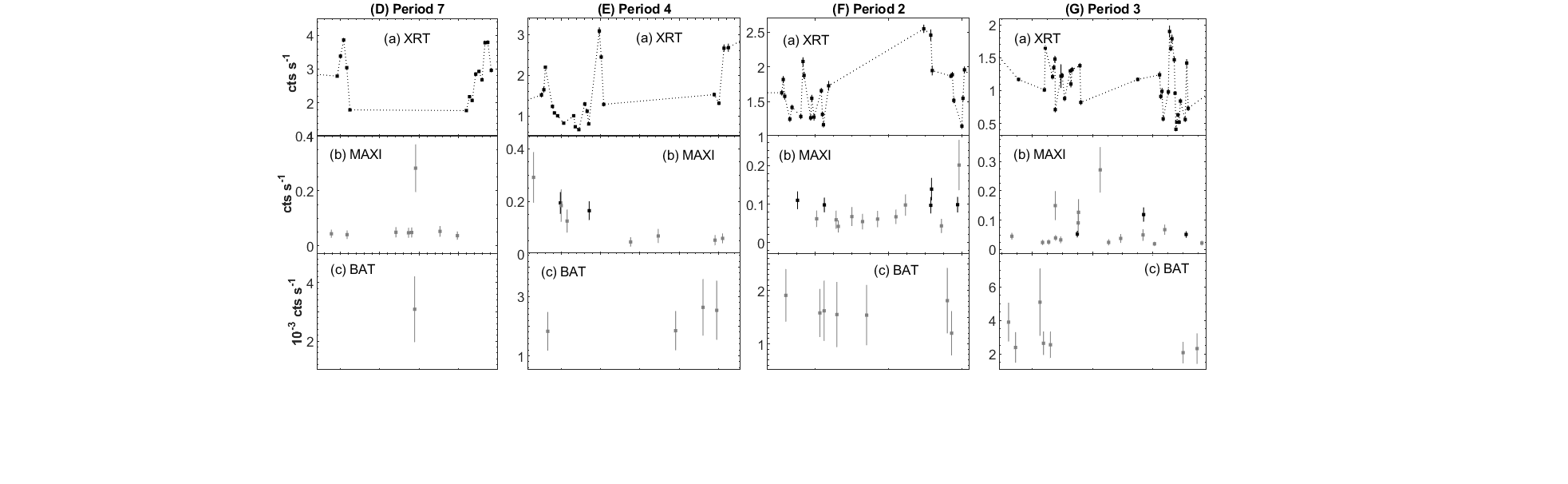}
\includegraphics[trim=5.9cm 1.6cm 0cm 0.33cm, clip=true, scale=0.88]{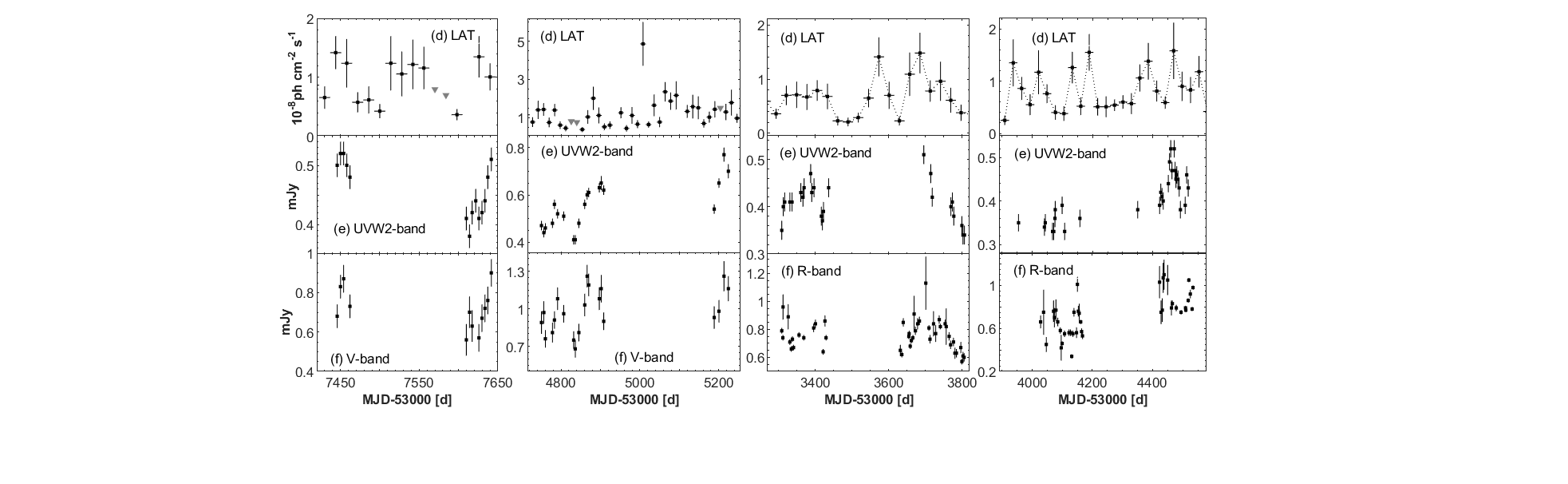}
\vspace{-0.5cm}
 \caption{\label{mwlper}  -- Continued. }  \vspace{-0.3cm}
 \end{figure*}

 Highest number of the detections with 5$\sigma$ significance (5 times) occurred in the case of 2-day-binned BAT observations and, therefore, this time integration was used for constructing the  15--150\,keV light curve in each period (see Fig.\,\ref{xrtper}). The first two detections preceded the start of the XRT observations of the target in 2005, and the last instance occurred in the epoch of the strong 0.3--10\,keV flare in 2012 March--May (Period\,1). The remaining two detections belong to the period 2009 February--2011 May (on MJD 54884 and 55665) when the source was not observed with XRT.   Similar to the MAXI-band observations, we also plotted the data points  corresponding to the detection with (3--4)$\sigma$  significances. The 2-day-binned BAT data also yielded the highest number of detections with these significances (91 times), which were relatively frequent in the epochs of the 5$\sigma$-detections and  0.3--10\,keV flares.

Fig.\,\ref{xrtper} demonstrates that, generally, 1ES\,1218$+$304 was not a bright $\gamma$-ray source during Periods 1--7 (as well as in other time intervals): the mean 0.3--300\,GeV flux during the entire period of our study amounted to $\sim$2$\times$10$^{-8}$\,ph cm$^{-2}$\,s$^{-1}$, and we used a 4-week integration for obtaining robust detections from all time bins, except for the three occasions from the period 2014 June 3--August 26 (MJD\,56811--56895; the interval between Periods 2 and 3). Otherwise, the robust detections amounted to  64\% from the two-week-binned data, whereas this occurred for less than 25\% in the case of the one-week integration. The shortest-term robust detection of the source occurred during the 1.01\,d time interval centered on MJD\,58059.5 (2017\,November\,2; Period\,4), characterized also by the highest historical 0.3--300\,GeV flux of (8.78$\pm$2.77)$\times$10$^{-8}$ph\,cm$^{-2}$s$^{-1}$.   The amplitude $F_{\rm var}$ from the entire data set was higher in the case of the two-week-integrated LAT-band observations (in spite to the increasing uncertainties) compared to that from the 4-week integration, and this quantity could be even higher since the source frequently was not robustly detectable in the lowest 0.3--300\,GeV states when adopting the 2-week integration. This result is due to the fast strong HE $\gamma$-ray flares in some periods (discussed in detail in Appendix\,A). The corresponding $F_{\rm var}$ values  were higher than those in the XRT-band in Periods\,1--2 and 4  (see Tables \ref{xrtper} and \ref{uvotper}). The source showed a flaring activity on various timescales and amplitudes in different epochs, which was stronger than those observed in the optical--UV energy range. The strongest LAT-band activity was recorded in 2020 (Period\,5), along with the highest optical--UV states. 

The source was observed 143--162 times in the UVOT bands (most frequently - in 
\emph{U} and \emph{UVW1}). {The variability summary in each band during Periods 1--7 are given in Table\,\ref{uvotper}. The source showed a much slower optical--UV variability compared the XRT and LAT bands (see the two bottom panels in Figs \ref{xrttable}A--\ref{xrttable}G)\footnote{Since the fluxes from the bands \emph{U}--\emph{UVW2} show very strong cross-correlations (as discussed in Section 5.2), we  plotted only the light curves for the highest and lowest-frequency bands}. Consequently, the $F_{\rm var}$ values from Periods 1--7 were significantly lower than their XRT and LAT-band "counterparts". The strongest optical--UV flares occurred around MJD 58490 and 58984, while the source showed  moderate 0.3--10\,keV states during the second instance. The source was observed sparsely with UVOT during the first flare: only 1--4 visits to  the source were performed in the  \emph{U}--\emph{UVW2} bands (versus 9 XRT visits in the same epoch), and no observations were carried out in the bands \emph{B}--\emph{V} at all. Meanwhile, the source showed the strongest outburst and  highest \emph{R}-band states (in the beginning of 2019\,January; see \citealt{myr19} and the quick-look look data published on the Tuorla website\footnote{https://users.utu.fi/kani/1m/1ES\_1218$+$304\_jy.html}). 
An alert of an even higher state was reported in \cite{j20}: the source showed $m_{\rm R}$=15.00 mag during the ATOM observations performed in 2020 April, coinciding with the highest historical fluxes in the bands \emph{U}--\emph{UVW2} (versus about 15.10\,mag during the aforementioned highest state).
 
 \begin{table*} \footnotesize  \tabcolsep 3.3pt \vspace{-0.3cm}
 \begin{minipage}{185mm}
\caption{\label{doubl} Summary of the 0.3--10\,keV flux doubling/halving instances. Col.\,(1): the ObsID(s) of the corresponding XRT observation(s); (2): the flux doubling/ halving timescale (days); (3) and (4): the initial and final flux values during the given instance (in cts\,s$^{-1}$)); (5): the figure number presenting the instance; (6)--(8):  the ranges of the photon indices ("PL" stands for a simple the power-law model; otherwise, the values of the photon index $a$ are provided), curvature parameter and  the  synchrotron SED peak position. } 
   \vspace{-0.1cm}
    \hspace{-0.8cm}
      \centering
   \begin{tabular}{ccccccccc}
  \hline
 ObsIDs&$\tau_{\rm d,h}$ (d)  & $F_i$ & $F_f$ &Fig. &$ a$ or $\Gamma$& $ b $ & $ E_{\rm p}$ (keV)  \\
   (1)& (2) & (3)  & (4)  & (5) & (6)  & (7)  & (8) \\
\hline
(3037600)7--8	&4.46	&0.67(0.05)	&1.71(0.05)&	2C	&1.97(0.05),  2.17(0.09)PL	&0.28(0.13)&	1.13(0.15)\\
(3037601)2--4	&4.21	&1.65(0.04)	&4.44(0.07)&	2C	&1.94(0.03)--1.99(0.04)PL&	-&	-\\
(3037604)4--5&	0.91	&1.52(0.05)	&0.69(0.04)&	2G	&1.97(0.05)L, 2.18(0.09)PL	&0.62(0.19)	&1.25(0.20)\\
(3037605)6--9&	10.21	&1.24(0.05)	&0.55(0.04)&	2G	&1.79(0.09), 1.82(0.09)—1.98(0.06)PL&	0.47(0.20)	&1.67(0.23)\\
(303760)59--61&	11.70	&0.55(0.04)&	1.90(0.09)&	2G	&1.69(0.07)—1.91(0.07)PL	&-&	-\\
(3037606)4--6&	3.13	&1.47(0.05)	&0.41(0.03)&	2G	&1.56(0.12), 
2.12(0.05)—2.21(0.14)PL	&1.06(0.37)&	1.610.39)\\
(3037607)1--2	&3.80	&0.56(0.04)&	1.42(0.06)&	2G	&1.66(0.08)L, 2.24(0.10)PL	&0.41(0.21)	&2.60(0.24)\\
(3037607)6—9&	20.92&	2.20(0.03)	&1.01(0.04)&	2G	&1.61(0.08)--1.76(0.08),
&	0.31(0.19)--0.58(0.20)	&2.17(0.22)--2.44(0.24)\\
&&&  & & 1.91(0.07)-- 2.07(0.06)PL	&  &     \\
(3037608)8--9&	13.94&	0.81(0.03)	&3.09(0.09)&	2E	&1.68(0.05)—2.13(0.06)P	&-	&-\\
(303760)89--92	&8.80	&3.09(0.09)&	1.29(0.05)&	2E	&1.82(0.07)—2.00(0.05),&
	0.28(0.12)--0.42(0.18)&	1.00(0.13)--1.64(0.19)\\
&&&  & & 1.68(0.05)—2.05(0.07)PL	&  &     \\
(3037612)7--8&	7.56	&2.55(0.05)	&1.13(0.04)&	2A	&1.63(0.09)—1.85(0.06),
1.84(0.05)PL&	0.42(0.15)--0.55(0.19)&	1.51(0.17)—2.17(0.21)\\
(303761)28--30	&13.21	&1.13(0.04)	&2.65(0.08)&	2A	&1.68(0.07)—1.85(0.06),
&0.24(0.13)--0.42(0.15)	&1.51(0.17)—4.64(0.41)\\
&&&  & & 1.71(0.08)—1.74(0.05)PL		&  &     \\
 \hline
\end{tabular}
\end{minipage}
\end{table*}

\subsection{Spectral results}
\subsubsection{X-ray spectral analysis}

The 0.3--10\,keV spectral analysis  was performed by using the \texttt{HEASOFT} package \texttt{XSPEC}  (version 12.14.1) and the latest response matrix from the \emph{Swift}-XRT calibration files. In order to use the $\chi^2$-statistics, the XRT instrumental channels were combined  to include at least 20\,photons per each bin (via the task \texttt{GRPPHA}). Each  spectrum was corrected for absorption by neutral hydrogen (both the H\,I and H$_{2}$ components) by adopting the corresponding column density fixed to the Galactic value of $2.06\times10^{20}$ cm$^{-2}$ \citep{w13}. First of all, the spectra were fitted with the log-parabola model  (developed by \citealt{m04}, and implemented via the XSPEC model \texttt{LOGPAR}):
\begin{equation}
F(E)=K(E/E_{\rm 1})^{-(a+b log(E/E_{\rm 1}))},
\end{equation}
with the reference energy $E_{\rm 1}$ fixed to 1\,keV; $a$,  the photon index at this energy; $b$, the curvature parameter;  $K$, the normalization factor. We checked the model validity  by means of the value of the reduced chi-squared ($\chi^2_{\rm r}$), as well as via the distribution of the fit residuals. When spectral curvature was detected with a significance of 2$\sigma$ or higher, the log-parabola model generally yielded better statistics compared to the spectral models generally adopted for blazars: broken\footnote{https://heasarc.gsfc.nasa.gov/xanadu/xspec/manual/node141.html} and simple  power laws, as well as log-parabola with low-energy power-law branch (LPPL; \citealt{m06,t09}). However, when the spectrum showed a curvature with a significance lower than 2$\sigma$, we re-fitted it using  (i) single power-law model $F(E)=KE^{-\Gamma}$, with $\Gamma$, the photon index throughout the entire  0.3--10\,keV energy range; (ii) the LPPL model.

 \begin{table*} \footnotesize  \tabcolsep 1.8pt   \vspace{-0.5cm}
 \begin{minipage}{180mm}
\caption{\label{idvtable} Summary of the 0.3--10\,keV IDVs (extract). Cols. (1)--(3) contain the ObsID(s) of the XRT observation(s) during which the given IDV was recorded ("Or" denotes "Orbit"), the corresponding MJD  and the total  length of the observation (in hours, including the intervals between the separate \emph{Swift} orbits), respectively; (4): reduced chi-squared and the corresponding degrees-of-freedom, along with the time bin used for the variability search (in seconds; "Or" stands for "Orbit"); (5): fractional variability amplitude; (6)--(8):  the ranges of the photon indices ("PL" denotes the power-law model; otherwise, the results are related to the photon index $a$), curvature parameter and  the position of the synchrotron SED peak. } 
   \vspace{-0.1cm}     \hspace{-0.8cm}       \centering
   \begin{tabular}{ccccccccccc}
  \hline
 ObsID& MJD & $\Delta$T\,(hr) & $\chi^2$/DOF/Bin &  $F_{\rm var}(\%)$  & $ a$ or $\Gamma$& $ b $ & $ E_{\rm p}$ (keV) \\
   (1)& (2) & (3)  & (4)  & (5) & (6)  & (7)  & (8)& & \\
 \hline
30376007&56000& 0.27 &4.57/7/120 & 21.6(4.2)& 2.17(0.09)\,PL& - &  -	\\
30376015&56066& 2.00 &3.56/32/120 & 3.6(0.8)&1.80(0.05)--1.88(0.05)   &0.22(0.09)--0.28(0.10)  & 1.64(0.14)--2.85(0.25) \\
&&  & &  &1.88(0.04)--1.92(0.03)\,PL  &  &  	\\
30376033 &56718& 0.23&3.81/6/120& 15.2(4.0)&1.60(0.07)  &0.32(0.17)   &4.22(0.36   \\
30376034   &56769&0.27  &3.51/7/120 & 9.3(2.2) &1.77(0.07), 1.89(0.05)\,PL &0.33(0.16)  & 2.23(0.23) 	\\
 \hline
\end{tabular}
\end{minipage}
\end{table*}

\begin{figure*} \vspace{-0.1cm}
\includegraphics[trim=6.1cm 5.6cm 0cm 0cm, clip=true, scale=0.9]{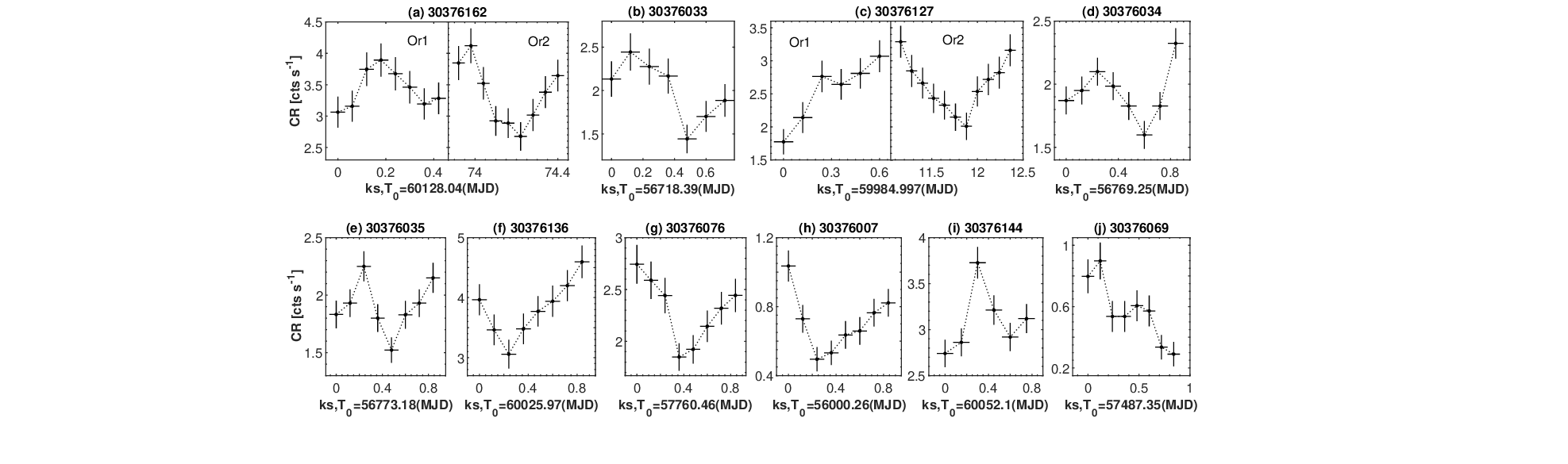}
\vspace{-0.6cm}
  \caption{\label{idv} Subhour 0.3–10 keV variability of 1ES\,1218$+$304 (extract). The IDV instances are presented in order of increasing exposure. }
 \end{figure*}

We performed an orbit-resolved spectral analysis when the specific XRT observation in the WT regime was distributed over two or more \emph{Swift} orbits. Moreover, we extracted spectra from some one-orbit XRT observations  when the source showed a 0.3--10\,keV IDV, or the fit residuals exhibited some trends and, therefore, no satisfactory fit with any of the aforementioned models were  achieved. This situation generally occurs when the spectral parameters (photon index, curvature parameter) vary fastly, or the  0.3--10\,keV emission undergoes a transition from the logparabolic into the power-law spectral distribution or conversely. For the same reasons, the spectra were extracted and analyzed from different segments of several XRT observations performed in the PC regime.

\begin{table*} \tabcolsep 2.5pt \small  \begin{minipage}{185mm} \vspace{-0.2cm} \hspace{-1cm}
  \caption{\label{uvotper} Same as Table\,\ref{xrtper} for the UVOT (1-d bins) and LAT (2\,weeks) observations. The optical--UV fluxes are given in mJy;  LAT-band photon flux - in 10$^{-8}$ph\,cm$^{-2}$s$^{-1}$.}
  \vspace{-0.2cm}   \centering  \hspace{-0.7cm}
  \begin{tabular}{ccccccccccccccccccc}
    \hline
     Per. &\multicolumn{3}{c}{$F_{\rm UVW2}$}&\multicolumn{3}{c}{$F_{\rm UVM2}$}  &  \multicolumn{3}{c}{$F_{\rm UVW1}$}    & \multicolumn{3}{c}{$F_{\rm U}$}   \\
  \hline
 &  Max  & Mean  & $F_{\rm var}(\%)$ & Max  & Mean   & $F_{\rm var}(\%)$ & Max   & Mean   & $F_{\rm var}(\%)$ & Max   & Mean   & $F_{\rm var}(\%)$    \\
(1) & (2) & (3) & (4)& (5)& (6) & (7) & (8)& (9)& (10)& (11)& (12)& (13)   \\
 \hline
1&0.78(0.02)&0.68(0.01)&6.6(1.1)&0.82(0.02)&0.73(0.01)&6.6(1.4)&0.85(0.03)&0.75(0.01)&7.0(1.5)&1.13(0.04)&0.96(0.01)&8.1(1.4)\\
2&0.51(0.02)&0.41(0.01)&9.1(1.1)&0.56(0.02)&0.45(0.01)&9.9(1.0)&0.49(0.03)&0.44(0.01)&7.0(1.5)&0.70(0.04)&0.59(0.01)&8.7(1.2)\\
3&0.52(0.02)&0.41(0.01)&12.8(0.9)&0.55(0.03)&0.44(0.01)&11.2(1.2)&0.60(0.03)&0.45(0.01)&14.2(1.1)&0.82(0.04)&0.60(0.01)&15.2(1.0)\\
4&0.77(0.03)&0.55(0.01)&17.5(0.9)&0.86(0.04)&0.60(0.01)&17.6(1.2)&0.86(0.0)&0.60(0.01)&17.0(1.1)&1.05(0.05)&0.78(0.01)&16.1(1.1)\\
5&1.56(0.05)&1.19(0.01)&21.7(0.9)&1.72(0.05)&1.34(0.01)&24.6(1.1)&1.75(0.06)&1.25(0.02)&33.5(1.5)&2.30(0.04)&1.90(0.02)&18.4(1.0)\\
6&0.95(0.03)&0.51(0.01)&18.1(0.7)&0.99(0.04)&0.56(0.01)&17.0(0.9)&0.97(0.05)&0.56(0.01)&16.3(0.9)&1.18(0.04)&0.71(0.01)&16.1(0.9)\\
7&0.52(0.02)&0.46(0.01)&9.3(1.2)&0.54(0.02)&0.49(0.01)&6.8(1.4)&0.58(0.03)&0.50(0.01)&7.5(1.7)&0.71(0.03)&0.63(0.01)&6.9(1.4)\\
 \hline
 & \multicolumn{3}{c}{$F_{\rm B}$} & \multicolumn{3}{c}{$F_{\rm V}$}&  \multicolumn{3}{c}{$F_{\rm LAT}$} & \multicolumn{3}{c}{$\Gamma_{\rm LAT}$} \\
    \hline
 &  Max  & Mean  & $F_{\rm var}$ (\%)& Max & Min& Mean  & Max   & Mean   & $F_{\rm var}(\%)$ & Max  & Min& Mean\\
(1) & (2) & (3) & (4)& (5)& (6) & (7) & (8)& (9)& (10)& (11) & (12)& (13) \\
 \hline
1&1.21(0.04)&1.08(0.02)&9.7(2.6.)&1.34(0.05)&1.11(0.03)&9.7(2.6)&3.14(0.61)&1.16(0.10)&58.4(10.1)&1.95(0.17)&1.23(0.13)&1.66(0.05)\\
2&0.76(0.04)&0.66(0.01)&-&0.82(0.07)&0.71(0.02)&-&3.32(0.98)&0.96(0.10)&53.6(8.5)&2.76(0.28)&1.33(0.15)&1.88(0.04)\\
3&0.89(0.07)&0.67(0.01)&14.2(2.3)&0.98(0.09)&0.74(0.02)&14.2(2.3)&2.45(0.59)&1.05(0.6)&33.8(7.6)&3.25(0.40)&1.17(0.16)&1.89(0.03)\\
4&1.11(0.05)&0.88(0.01)&15.1(2.1)&1.26(0.12)&0.98(0.02)&15.1(2.1)&4.86(1.15)&1.27(0.06)&56.1(6.6)&2.90(0.27)&1.26(0.17)&1.81(0.03)\\
5&2.82(0.07)&2.11(0.03)&24.1(2.1)&2.93(0.11)&2.34(0.05)&24.1(2.1)&4.45(1.05)&1.48(0.06)&44.0(5.0)&2.58(0.24)&1.47(0.14)&1.82(0.03)\\
6&1.21(0.05)&0.79(0.01)&16.2(1.9)&1.37(0.10)&0.78(0.02)&16.2(1.9)&3.06(1.12)&1.02(0.05)&45.0(6.8)&2.63(0.29)&1.31(0.15)&1.83(0.03)\\
7&0.90(0.07)&0.72(0.02)&10.6(3.2)&0.85(0.04)&0.72(0.01)&11.5(3.3)&1.42(0.28)&0.95(0.08)&28.2(9.4)&2.18(0.20)&1.23(0.15)&1.70(0.04)\\
 \hline \end{tabular} \end{minipage} \end{table*}

For curved spectra, the position of the synchrotron SED peak was calculated as $E_{\rm p}=10^{(2-a)/2b}$\,keV (\citealt{m04} and references therein). Note that the practically  same result is obtained by the XSPEC model \texttt{EPLOGPAR} presenting the spectrum as  $F(E)=K_{\rm s}10^{-b(\log(E/E_{\rm p}))^2}$\,keV ($K_{\rm s}$, the norm; \citealt{t07}): the difference
from the value calculated by using \texttt{LOGPAR} is significatly lower compared to the error range. Moreover, these two models also yield practically the same values of the parameter $b$. Finally, we determined the unabsorbed 0.3--2\,keV, 2--10\,keV, and 0.3--10\,keV fluxes by using the tool \texttt{EDITMOD}. The soft 0.3--2\,keV and hard 2--10\,keV fluxes showed   a strong cross-correlation (see Table\,\ref{cortable}). However, the range of 
the hard X-ray flux values was twice wider, the maximum flux about 40\,per cent higher and fractional variability amplitude larger than in the case of the soft flux.

\subsubsection{Properties and timing of  spectral parameters}

About 49\,per cent of the 0.3--10\,keV spectra (126 out of a total of 258) show a significant curvature (see Table\,\ref{logp} for the corresponding results). The distribution of values of the parameter $b$ (corresponding to the curvature detection with the significance of 3$\sigma$ and higher) is presented in Fig.\,\ref{figdistr}a and the corresponding properties (minimum, maximum, mean  and peak values, distribution skewness) are presented in Table\,\ref{distrtable}. Note that 1ES\,1218$+$304  frequently showed large curvatures ($b$$\sim$0.5 or higher). Note that this parameter showed a variability on diverse timescales,  and the spectra with a relatively low curvature were mainly observed during the higher X-ray states. Consequently, the source showed anti-correlation between the curvature and de-absorbed 0.3--10\,keV flux (weak but statistically significant; Fig.\,\ref{figcor}b and Table\,\ref{cortable})\footnote{We used only those values of the parameter $b$ for the correlation study, which correspond to the curvature detection with the significance of 3$\sigma$ and higher.}.  The most extreme variability occurred during the 0.3--10\,keV IDV recorded on MJD\,58500 (see Tables \ref{idvtable} and \ref{logp}): the curvature initially increased by $\Delta b$=0.32(0.17)  within 6.3\,ks and then dropped by $\Delta b$=0.31(0.18)  during the subsequent 0.6\,ks interval. On the contrary, the curvature dropped with $\Delta b$=0.30(0.19)  in 39.7\,ks and increased with $\Delta b$=0.26(0.18)  within the subsequent 6.6\,ks interval during another IDV, which occurred on MJD\,58498.

Note also that the  curved spectra contain two subsets of different properties: (1) "Subset-1", characterized by the negative trend $b$--$E_{\rm p}$ and positive $a$--$b$ correlation (see Table\,\ref{cortable} and Figs \ref{figcor}c--\ref{figcor}d); (2) "Subset-2", notable for the anti-correlation between the parameters $a$ and $b$ (Fig.\,\ref{figcor}e). Note that the  $b$--$E_{\rm p}$ trend is not observed for the latter subset. 

The photon index at 1\,keV was characterized by a broad range of values  $\Delta a$=0.93$\pm$0.1 and the most extremely hard spectrum well-fit with $a_{\rm min}$=1.53$\pm$0.08 (see Fig.\,\ref{figdistr}b). Fig.\,\ref{figdistr}b demonstrates that a majority of the curved spectra were hard ($a$$<$2) or  very hard ($\Gamma$$<$1.80). On average, the source showed the hardest spectra during Period\,6 (the mean value $\overline{a}$=1.82$\pm$0.01), contrary to the softest spectra with $\overline{a}$=2.02$\pm$0.02 in Period\,1 (see Table\,\ref{xrtper}).
The photon index varied by $\Delta a$=0.10(0.06)--0.21(0.07)
during the several 0.3--10\,keV IDVs, as well as  some multi-segment observations without IDVs (see Tables \ref{idvtable} and \ref{logp}). Namely, the largest intra-day spectral variability  (a softening) occurred on MJD\,60454  within 1.2\,ks exposure. While a drop in the 0.3--10\,keV flux is expected in such a situation, the spectral curvature showed a decrease (i.e., a potential brightening) and, eventually, no intraday variability occurred. 
During the XRT-band flares, the largest variabilities of the photon index were a hardening by $\Delta a$=0.40(0.10) and the subsequent softening by $\Delta a$=0.36(0.09) during MJD\,(589)82--96 (Period 5). 

In the course of these changes, the source mainly followed a  "harder-when-brighter" (HWB) spectral trend, reflected in the anti-correlation $a$--$F_{\rm 2-10 keV}$ (Fig.\,\ref{figcor}f and Table\,\ref{cortable}). However, the latter was not strong, and the corresponding scatter plot shows outliers from the general trend, as well as a bend after the level $F_{2-10 keV}$$\approx$4$\times$10$^{-11}$\,erg\,cm$^{-2}$\,cm$^{-1}$: the data set above this threshold is characterized by a significantly lower slope compared to the lower-energy set. No clear spectral trend or even the opposite ("softer-when-brighter") evolution was observed during some time intervals of various durations: from intraday timescales to about two months (during MJD\,56330--56390, although incorporating only four observations and, therefore, no firm conclusion can be drawn).  Consequently, some very hard spectra do not correspond to the highest 2--10\,keV states of the source. Finally, the softest spectra also made an outlier from the scatter plot $a$--$F_{\rm 2-10 keV}$.

The  synchrotron SED peak position also showed a wide range  of values between $E_{\rm p}$=0.21$\pm$0.31\,keV and $E_{\rm p}$=4.83$\pm$0.42\,keV (see Tables \ref{logp} and \ref{distrtable}). Note that  this peak was poorly constrained by the XRT observations when $E_{\rm p}$$\lesssim$0.5\,keV (4 spectra). Consequently,  the corresponding $E_{\rm p}$ values represent upper limits to the intrinsic SED peak position (see, e.g., \citealt{k20,k23b}), and the corresponding values are not included in Fig.\,\ref{figdistr}c. Note that the fits of these spectra with the LPPL model (both in the cases of only XRT data or including also those from the simultaneous UVOT observations) did not yield better statistics and  curvature detection at the significance of 2$\sigma$ and higher.

The parameter $E_{\rm p}$ varied on diverse timescales and was showing an intraday variability five times (see the last column of Table\,\ref{idvtable}). The most extreme variability occurred on MJD\,56066 when the SED peak position was shifted by 1.21$\pm$0.29\,keV towards lower energies within 2\,hr. On longer timescales, the largest changes during the densely sampled observations were the shifts by 3.67$\pm$0.44\,keV towards the lower energies (during MJD\,(5603)4--9)  and by  3.13$\pm$0.43\,keV to the opposite direction (MJD\,(606)29--33).  Predominantly, the source exhibited a shifting the synchrotron SED peak to higher energies with the increasing X-ray flux, yielding a positive correlation between  $E_{\rm p}$ and unabsorbed 2--10\,keV flux (see Fig.\,\ref{figcor}g and Table\,\ref{cortable}). Moreover, a positive correlation between $\log E_{\rm p}$ and $\log S_{\rm p}$ was also observed (see Fig.\,\ref{figcor}h and Section 5.1.2 for the corresponding physical implication), where the height of the synchrotron SED peak $S_{\rm p}$ was calculated 
as $S_{\rm p}$=1.6$\times$10$^{-9}\,K$10\,$^{(2-a)^2/4b}$\,erg\,cm$^{-2}$s$^{-1}$ (\citealt{m04} and references therein).

As noted above, more than 51\% of all spectra  did not show a significant curvature and was well-fit with a simple power-law model (see Table\,\ref{pow}). Note that the use of the LPPL model did not yield a curvature detection with a significance of 2$\sigma$ and higher also for these spectra. 
The range of the derived photon index $\Gamma$  was wider compared to that from the curved spectra ($\Delta \Gamma$=1.09(0.09), $\Gamma_{\rm min}$=1.60$\pm$0.06). Fig.\,\ref{figdistr}d demonstrates that  the higher fraction of the power-law spectra was characterized by $\Gamma$$ \geqslant$2 (at least 27\,per cent, taking into account the error ranges) compared to the curved spectra (10\,per cent).  The power-law spectra were present in all periods, but their portion was the highest during Periods\,3--4 (67--69\,per cent of all spectra; see Table\,\ref{xrtper}).  The source frequently showed a transition from a logparabolic SED into a power-law one (or/and conversely) even within a single XRT observation. Such a situation is presented in Fig.\,\ref{recon}: the source showed a power-law SED from the spectra extracted from the three 700-second segments of the first orbit and from the 600-second  segment of the second orbit (with $\Gamma$=1.88(0.04)--1.92(0.03)) of ObsID\,30376015, while the first and third segments of the latter were characterized by the logparabolic spectra with $a$=1.80(0.05)--1.88(0.05), $b$=0.22(0.09)--0.28(0.10) and $E_{\rm p}$=1.64(0.15)--2.85(0.25) keV.

\begin{table*}
\vspace{-0.3cm}
 \tabcolsep 4pt \centering \footnotesize
\begin{minipage}{170mm}
\caption{\label{logp} Results of the 0.3--10\,keV spectral analysis with the log-parabolic model (extract). In col.\,(1), The acronyms "Or" and  "S" stand for  "Orbit" and "Segment", respectively.  The $E_{\rm p}$ values (col.\,4) are given in keV. Col.\,(6): the reduced Chi-squared along with the corresponding.  The unabsorbed 0.3--2\,keV, 2--10\,keV and 0.3--10\,keV fluxes (Cols.\,7--9) are provided in  10$^{-11}$erg\,cm$^{-2}$s$^{-1}$.} \vspace{-0.1cm}
\centering  \begin{tabular}{ccccccccc}     \hline
ObsId & $a$ & $b$ &$E_{\rm p}$  & 100$\times K$ & $\chi^2$/DOF & $F_{\rm 0.3-2\,keV}$ & $ F_{\rm 2-10\,keV}$ & $ F_{\rm 0.3-10\,keV}$  \\
(1) & (2) & (3) & (4) & (5) & (6) & (7) & (8) & (9)  \\
\hline  
 35016001&	2.09(0.04)	&0.16(0.08)&	0.52(0.09)&	9.15(0.22)&	1.12/102	&2.82(0.07)	&1.73(0.12)	&4.56(0.14)\\
30376001	&2.18(0.05)	&0.21(0.11)	&0.37(0.12)	&5.16(0.19)	&0.88/52&	1.58(0.05)&	0.84(0.08)&	2.43(0.10)\\
40578002	&2.13(0.05)	&0.34(0.13)&	0.64(0.14)&	9.91(0.35)	&1.00/62	&2.95(0.10)	&1.52(0.16)	&4.47(0.19)\\
30376008	&1.97(0.05)	&0.28(0.13)&	1.13(0.15)&	10.44(0.38)	&0.97/57	&3.02(0.11)	&2.12(0.20)&	5.14(0.23)\\
\hline \end{tabular} \vspace{0.4cm} \end{minipage}
\end{table*}

  \begin{figure*}
\vspace{-0.6cm}
\includegraphics[trim=6.1cm 5.6cm -1cm 0cm, clip=true, scale=0.91]{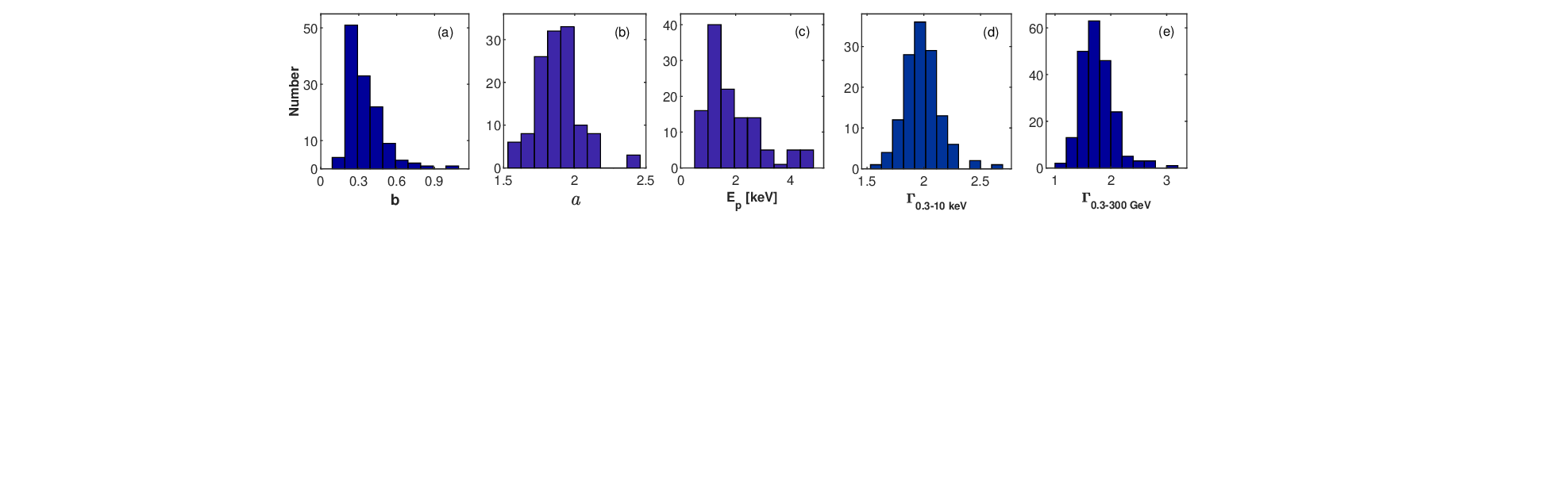}
\vspace{-0.8cm}
 \caption{\label{figdistr} Distribution of different spectral parameters. The LAT-band photon-indices to the 4-weekly-integrated data.}
 \vspace{-0.2cm}
 \end{figure*}

The power-law spectra also showed a HWB trend, with the hardest and softest spectra corresponding to the highest and lowest X-ray states, respectively. However, there were the sub-samples characterized by the different slopes in the flux--photon index plane, as well as  some time intervals showing no clear spectral trend or even a  "softer-when-brighter" spectral evolution (from intraday timescales to several days). Consequently, the anti-correlation  $\Gamma$--$F_{\rm 2-10 keV}$ was not strong, similar to the curved spectra (see Table\,\ref{cortable}). The photon index varied  by $\Delta a$=0.09(0.06)--0.18(0.06) during the several 0.3--10\,keV IDVs and multi-segment observations (Tables \ref{idvtable} and \ref{logp}). The largest intra-day spectral hardening occurred on MJD\,58492 (MJD\,30376172) within the 7.8-ks interval.

Finally, the 0.3--300\,GeV photon index from the four-week-binned LAT observations (derived via the power-law fit to the HE spectrum) also showed a wide range $\Delta \Gamma$=2.20$\pm$0.42, with the hardest value $\Gamma_{\rm min}$=1.05$\pm$0.18 (Table\,\ref{distrtable}). The corresponding distribution show a dominance of the hard and very hard spectra with the peak value $\Gamma_{\rm p}$=1.68(0.02) (Fig.\,\ref{figdistr}e). Moreover, this histogram shows the presence of the extremely hard spectra with $\Gamma$$\lesssim$1.5. Note that the source showed a softer-when-brighter  trend in this energy range (Fig.\,\ref{figcor}j), which is opposite to that demonstrated in X-rays. The fastest and largest subsequent hardening and softening were $\Delta \Gamma$=-1.82(0.41) and  $\Delta \Gamma$=1.62(0.31) each within 4 weeks, respectively. In attempt to check whether the higher-energy SED component was peaking  in the 0.3--300\,GeV band during some intervals during the period of our study, we fitted the spectra extracted from the 4-week binned LAT data with the log-parabola model.  However, the target's faintness and possible presence of the $\gamma$-ray SED peak did not allowed to detect the spectral curvature with a significance of 2$\sigma$ and higher for the 4-week-binned data. This was possible for significantly wider time intervals, characterized by softening of the HE spectrum beyond 10\,GeV and explained by possible significant contribution of from the photons energized via the IC-upscatter of X-ray emission in the Klein-Nishina (KN) regime (see Table\,\ref{klein} and Section\,5.2.2 for the corresponding discussion). The spectra were characterized by $b$=0.13(0.06)--0.20(0.06) and showed the presence of the $\gamma$-ray SED peak beyond 100\,GeV. The latter was calculated as $E_{\rm p}=E_0 10^{(2-a)/2b}$\,GeV, with $E_0$ to be the reference energy;  $a$, the photon index at  $E_0$. The latter showed the range $a$=0.94(0.21)--1.37(0.18).

 \section{Discussion and conclusions}

\subsection{particle acceleration and emission mechanisms}

As demonstrated by our detailed X-ray study of 1ES\,1218$+$304, the source underwent a  strong variability of the spectral properties over the 19-year period of the \emph{Swift} observations. Namely, it showed a wide range of the synchrotron SED position (up to 5\,keV) and always was an X-ray peaked HBL, even in the case of the spectra yielding $E_{\rm p}$$<$0.5\,keV: we  estimated the intrinsic SED peak position by fitting their broadband synchrotron SED with the log-parabolic function $log\nu F_{\rm \nu} = A(log \nu)^2 + B(log \nu) + C$ \citep{l86}. During the most XRT observations, our target was an extreme high-frequency peaked object (EHBL) with $E_{\rm p}$$>$1\,keV, according to the classification of \cite{c01}. Note also that all those $E_{\rm p}$ values reported in the framework of the previous studies are included in the range presented in Table\,\ref{distrtable} (see, e.g., \citealt{m08,s08,c18}).

Our target is among 13 HBL sources showing $a$$\sim$1.5 (see \citealt{k22b} for the corresponding review).  On the other hand, the particle acceleration processes in the target's jet were probably relatively less powerful compared to  six objects of this sample (1ES\,0033$+$595, Mrk\,421, Mrk\,501, H1515$+$660, 1ES 2344$+$514, BZB\,J1137$-$1710): the latter sometimes showed even significantly harder spectra (down to $a$$\lesssim$1 in the case of 1ES\,0033$+$595), and the synchrotron SED peak sometimes shifted beyond 10\,keV. Moreover, tens of the power-law spectra were hard or very hard down to $a$$\sim$1.6, and even harder values of the index $\Gamma$ have been reported only for 4 HBLs (Mrk\,501, 1ES\,0229$+$200, BZB\,J0832$+$3300 and PKS\,0548$-$322 (taking into account the associated errors; \citealt{k22b}). Nevertheless, the ranges of the photon indices shown by the target ($\Delta a$$\sim$=0.9 and $\Delta \Gamma$$\sim$1.1) during the 19\,yr period, very fast variability of the different spectral parameters (on subhour timescales), transition from the logparabolic electron energy distribution (EED) into power-law one and/or  conversely within a single XRT exposure hints at the extreme variability of the physical condition in the target's jet.  

\subsubsection{Fermi-type accelerations}

1ES\,1218$+$304 frequently showed  larger spectral curvatures ($b$$\sim$0.5 and higher) compared to some TeV-detected HBLs (e.g., M4k\,421 and Mrk\,501; see \citealt{k23b,k24b}). This result and the presence of the $b$-distribution peak at $b$=0.34 allows us to draw some conclusions about the acceleration processes operating in the target's jet. Namely,  a curved EED (that produces a logparabolic photon distribution with energy) can be established by a stochastic (or second-order Fermi) acceleration in
the highly turbulent jet medium  \citep{t09}, and  the synchrotron SED is expected to be relatively broad (i.e., spectral curvature with $b\sim0.3$ or lower) when this acceleration is  efficient \citep{m11}. On the other hand, a significant $b$--$E_{\rm p}$ trend is expected in that case  and has been reported from X-ray spectral studies of different HBLs (e.g., \citealt{m08,k20,k23b}). Note that this trend was not observed for the entire set of the spectra showing a curvature with 3$\sigma$ significance and higher. However, it was detected after removing the subset  showing the anti-correlation $a$--$b$  (denoted as Subset-2; see Section\,4.2.2). The latter can be established in the framework of the so-called energy-dependent acceleration probability (EDAP) process from the electron population characterized by a very low initial energy $\gamma_0$ \citep{m04,k20,k22b}. In such a situation, one can observe very hard  values of the photon index at 1\,keV: the EDAP-process (the specific case of first-order Fermi mechanism) may accelerate the aforementioned electron population mainly to the energies required to emit synchrotron  photons around $E$$\sim$1\,keV and much less frequently to the higher energies  (see \citealt{k22b} for the similar result in the case of 1ES\,0033$+$595). Even the X-ray emission zone contained a population of higher-energy electrons during those XRT observations, a strong cooling and short escape timescale (see below) could suppress the emission at hard X-ray frequencies. Consequently, the produced synchrotron SED should be characterized  by  a large spectral curvature and hard photon index at 1\,keV, which represents the reference energy for the logparabolic model. Note that these features are inherent to most spectra of the subset without the negative trend $b$--$E_{\rm p}$ and  showing the anti-correlation $a$--$b$. Since the latter can be  characterized by the comparable acceleration and cooling timescales, the aforementioned trend approaches asymptotic values and no significant $b$--$E_{\rm p}$ trend is observed. 

 \begin{table}
 \centering
  \begin{minipage}{85mm}
  \caption{\label{cortable} Correlations between different spectral parameters and MWL fluxes. Cols. 2 and 3 present the Spearman's correlation coefficient and the corresponding $p$-chance, respectively.}
  \vspace{-0.2cm}
  \begin{tabular}{ccc}
  \hline
  Quantities & $\rho$  & $p$ \\
 (1) & (2) & (3)   \\ 
  \hline
$F_{\rm 0.3-2\,keV}$ and $F_{\rm 2-10\,keV}$ & 0.88(0.03) & $<10^{-15}$ \\  
$b$ and $F_{\rm 0.3-10\,keV}$ & -0.38(0.11) & $9.34\times10^{-5}$ \\
$b$ and $E_{\rm p}$ (Subset-1) & -0.49(0.12) & $9.34\times10^{-5}$ \\
$b$ and $a$ (Subset-1) & -0.62(0.12) & $1.30\times10^{-7}$ \\
$b$ and $a$ (Subset-2) & -0.52(0.12) & $8.90\times10^{-6}$ \\
$a$ and $F_{\rm 2-10\,keV}$ & -0.49(0.09) & $5.52\times10^{-11}$ \\
$E_{\rm p}$ and $F_{\rm 2-10\,keV}$ & 0.62(0.10) & $8.82\times10^{-8}$ \\
$\Gamma$ and $F_{\rm 2-10\,keV}$ &-0.56(0.07) & $6.90\times10^{-12}$ \\
$\log E_{\rm p}$ and $\log S_{\rm p}$ & 0.46(0.15) & $3.22\times10^{-5}$  \\
$F_{\rm 0.3-2\,keV}$ and $F_{\rm UVW2}$ & 0.37(0.12) & $2.04\times10^{-6}$  \\ 
$F_{\rm 0.3-2\,keV}$ and $F_{\rm UVW1}$ & 0.33(0.12) & $6.42\times10^{-5}$  \\ 
$F_{\rm 0.3-2\,keV}$ and $F_{\rm U}$ & 0.28(0.12) & $3.57\times10^{-4}$  \\
$F_{\rm 0.3-2\,keV}$ and $F_{\rm V}$ & 0.24(0.12) & $1.74\times10^{-3}$  \\
$E_{\rm 0.3-300\,GeV}$ and $F_{\rm 2-10\,keV}$ & 0.34(0.11) & $5.85\times10^{-4}$ \\
$E_{\rm 0.3-300\,GeV}$ and $F_{\rm UVW2}$ & 0.41(0.11) & $6.02\times10^{-5}$ \\
$E_{\rm 0.3-300\,GeV}$ and $F_{\rm UVM2}$ & 0.40(0.11) & $1.95\times10^{-5}$ \\
$E_{\rm 0.3-300\,GeV}$ and $F_{\rm UVW2}$ & 0.36(0.12) & $8.84\times10^{-4}$ \\
$E_{\rm 0.3-300\,GeV}$ and $F_{\rm U}$ & 0.33(0.12) & $2.14\times10^{-4}$ \\
$E_{\rm 0.3-300\,GeV}$ and $F_{\rm B}$ & 0.31(0.12) & $7.22\times10^{-3}$ \\
$F_{\rm UVW2}$ and $F_{\rm UVW1}$ & 0.97(0.01) & $<10^{-15}$  \\
$F_{\rm UVW2}$ and $F_{\rm U}$ & 0.94(0.02) & $<10^{-15}$  \\
$F_{\rm UVW2}$ and $F_{\rm V}$ & 0.89(0.03) & $<10^{-15}$  \\
$F_{\rm UVW2}$ and $F_{\rm R}$ & 0.86(0.05) & $1.44\times10^{-13}$ \\
$m_{\rm UVW2}$ and $m_{\rm V}-m_{\rm UVW2}$& 0.41(0.10) &1.8$\times10^{-7}$  \\
$m_{\rm UVW2}$ and $m_{\rm UVW2}-m_{\rm B}$& 0.34(0.12) &9.2$\times10^{-5}$  \\
$m_{\rm B}$ and $m_{\rm B}-m_{\rm V}$& 0.31(0.12) &2.4$\times10^{-3}$  \\
$F_{\rm var}$ and $F_{\rm 0.3--10 keV}$ & -0.53(0.10) & 2.0$\times10^{-6}$  \\
$\Gamma$ and $F_{\rm 0.3-300\,GeV}$ &0.44(0.11) & $3.66\times10^{-6}$ \\
$F_{\rm 0.3-10\,GeV}$ and $F_{\rm 10-300\,GeV}$ &0.50(0.10) & $1.10\times10^{-6}$ \\
  \hline \end{tabular} \end{minipage}   \end{table}

On the other hand, the extremely hard values of the photon index $a$ can be obtained in the framework of the time-dependent stochastic acceleration scenario, depending on the balance between acceleration and escape times \citep{b06}. Namely, the slope of the spectrum produced by the stochastic acceleration is given by $n\simeq 1+t_{\rm acc}/2t_{\rm esc}$ where $t_{\rm acc}$ and $t_{\rm esc}$ represent the particle acceleration and escape timescales, respectively  \citep{k06}. The resulted curved EED can be extremely hard in the case of "hard-sphere" turbulence (with the spectrum $W(k)$$\propto$$k^{-q}$, where $k$ is the wave number; $q$=2) and specific balance between the acceleration and escape timescales(e.g., in the case of significantly longer escape timescales; \citealt{b06}). However, no anti-correlation between $a$ and $b$ parameters (i.e., large curvatures along with the very and extremely hard photon index, inherent for the subset without the negative trend $b$--$E_{\rm p}$) is predicted in that case.

\begin{table}
\tabcolsep 3pt
 \centering 
  \begin{minipage}{80mm}
  \caption{\label{distrtable} Distribution of different spectral parameters: the minimum, maximum, mean and peak values (cols. 2--5); distribution skewness (col. 6).}  \vspace{-0.1cm}
   \begin{tabular}{ccccccc}
  \hline
Par. & Min. & Max.  & Mean  & Peak& Skew. \\
(1) & (2) & (3) & (4) & (5) & (6) \\
\hline
$b$ &0.16(0.08) & 1.06(0.37) & 0.34(0.01) &0.24(0.02) &1.84 \\
$a$ &1.53(0.08)  &2.46(0.07)  &1.90(0.01) & 1.95(0.03) & 0.46\\
$\Gamma_{\rm XRT}$ &1.60(0.06) &2.69(0.07)  &1.98(0.01)  &1.95(0.02) & 0.27  \\
$E_{\rm p}$ &0.21(0.13)  &4.83(0.42)  &1.88(0.02) &1.25(0.04)  &1.25 \\
$\Gamma_{\rm LAT}$ &1.05(0.18) &3.15(0.38)  &1.75(0.01)  &1.68(0.02) & 0.49  \\
\hline \end{tabular} \end{minipage} \end{table}

Otherwise (i.e., when the initial electron energy is not very low), the EDAP process should yield a positive $a-b$ correlation  \citep{m04,m06}, which was the case for the subset of the curved spectra showing this correlation, along with the frequent occurrence of $b$$\sim$0.3 and showing  a negative trend $b$--$E_{\rm p}$ (inherent to the stochastic acceleration; see, e.g., \citealt{t09,t11,m11} ).  Note that the EDAP mechanism is not restricted by low $b$-values (in accord to the presence of those large $b$-values derived for some spectra of the subset with the positive $a-b$ correlation; see Fig.\,\ref{figcor}d) and could be characterized by a different slope of the scatter plot  $b$--$E_{\rm p}$ than the stochastic acceleration. Eventually, a joint operation of these two mechanisms could yield a significant scatter in the scatter plot and weaken the negative  $b$--$E_{\rm p}$ trend (see also Table\,\ref{cortable}).

\begin{figure*}
\vspace{-0.1cm}
\includegraphics[trim=6.0cm 0.5cm 0cm 0cm, clip=true, scale=0.91]{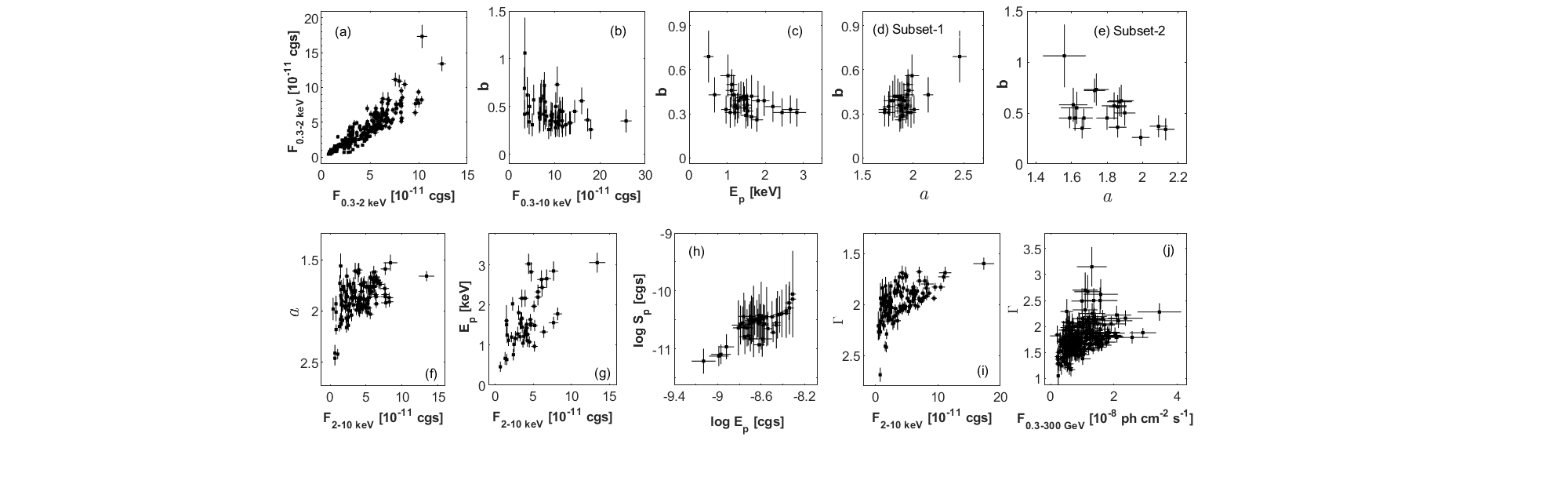}
\includegraphics[trim=5.8cm  0.8cm 0cm 0cm, clip=true, scale=0.90]{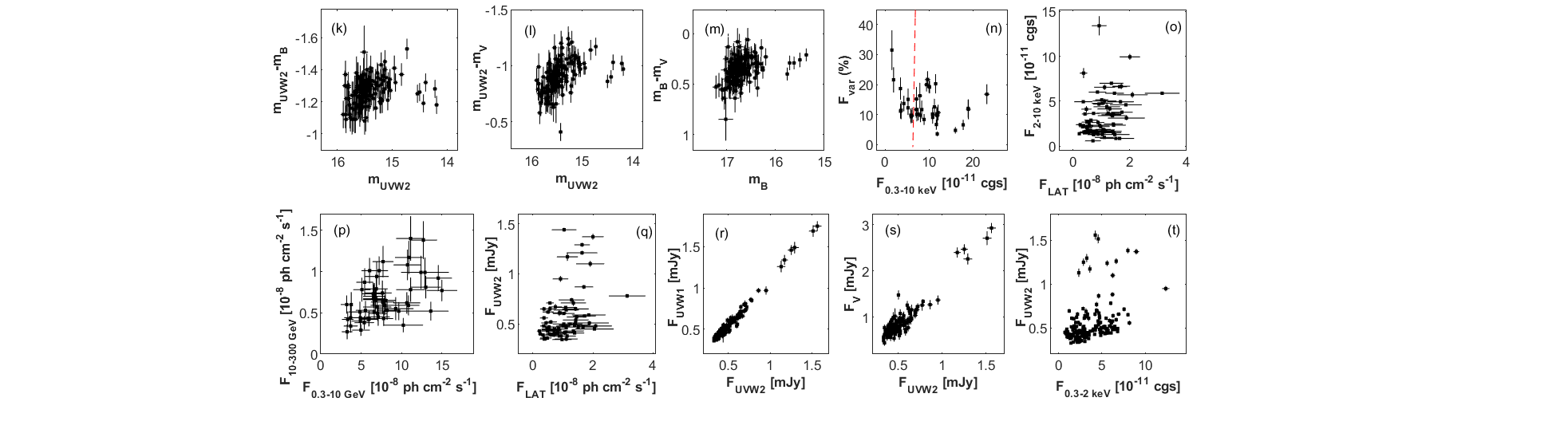}
\vspace{-0.4cm}
 \caption{\label{figcor} Correlations between the different spectral parameters, as well as  between spectral parameters and MWL fluxes. Panel\,(a): unabsorbed 0.3--2\,keV  and 2--10\,keV fluxes. The acronym "cgs" stands for erg\,cm$^{-2}$s$^{-1}$; (b)--(e): curvature parameter \emph{b} versus the unabsorbed 0.3--10\,keV flux, the position of the synchrotron SED peak $E_{\rm p}$, and the photon index \emph{a};  (f)--(g): the parameters \emph{a} and $E_{\rm p}$  plotted versus the 2--10\,keV flux; (h): the scatter plot $\log E_{\rm p}$--$\log S_{\rm p}$; (i)--(j): the 0.3--10\,keV and 0.3--300\,GeV photon indices versus the 2--10\,keV and 0.3--300\,GeV fluxes (1-week bins), respectively;  (k)--(l): the \emph{UVW2}-band magnitude  versus the colour-indices $m_{\rm UVW2}-m_{\rm B}$ and $m_{\rm UVW2}-m_{\rm V}$, and the same for  \emph{B}-band magnitude vs $m_{\rm B}-m_{\rm V}$ index is provided in panel\,(m); (n): fractional variability amplitude 0.3--10\,keV IDVs versus the corresponding unabsorbed flux (the vertical dashed red line stands for me mean 0.3--10\,keV flux for the entire period 2005--2024); (o) and (q): the  0.3--300\,GeV flux versus the 2--10\,keV and \emph{UVW2}-band  fluxes, respectively; (p): eight-weekly-binned 0.3--10\,GeV and 10--300\,GeV photon fluxes; (r)--(s):  \emph{UVW2}-band flux versus the \emph{UVM2} and \emph{V}-band fluxes, respectively; (t): unabsorbed 0.3--2\,keV flux versus the \emph{UVW2}-band flux (1\,d bins).}
  \end{figure*}
  
A relatively frequent observation of large spectral curvatures could be one of the reasons of the significantly lower rate of the target's TeV-detection than in the case of M4k\,421 or Mrk\,501, which generally show lower curvatures (along with a large difference in distance): a low probability of producing the TeV photons is predicted for such curvatures \cite{m11}. 

As noted in Section\,4.2.2, the source showed a positive correlation $\log E_{\rm p}$--$\log S_{\rm p}$, and the slope of the corresponding scatter plot (Fig.\,\ref{figcor}h) is $\alpha$=1.36$\pm$0.27. This result is close to the case the source showed the relation $S_{\rm p}$$ \varpropto$$E^{\alpha}_{\rm p}$ with $\alpha$=1.5. The latter is expected when the spectral changes are dominated  by variations in the average  electron energy and the number of X-ray emitting particles $N$ is also variable \citep{t09}. Note that both these conditions could be related to all above-discussed acceleration mechanisms. Note that a similar result was presented by \cite{w19} based on the X-ray observations of 1ES\,1218$+$304 performed with the different X-ray instruments during 1999-2017. However, different HBLs (Mrk\,421, Mrk\,501, 1ES\,1959$+$650) showed $\alpha$$\sim$0.6, expected by transitions in the turbulence spectrum during X-ray flares \citep{t11,k23b,k24b}. Moreover,  $\alpha$$\sim$1 for 1ES\,0033$+$595 during 2005--2022, which is expected when  spectral variability is dominated only by variations in the average electron energy \citep{t09,k22b}.

\begin{table*} \tabcolsep 4pt  \centering \footnotesize
    \begin{minipage}{140mm}
  \caption{\label{pow} Results of the XRT spectral analysis with a simple power-law model (extract). The unabsorbed 0.3--2\,keV, 2--10\,keV and 0.3--10\,keV fluxes (Cols.\,5--7) are presented in  10$^{-11}$erg\,cm$^{-2}$s$^{-1}$.}
     \vspace{-0.1cm}     \centering
   \begin{tabular}{ccccccc}
\hline
ObsId & $\Gamma$ & 100$\times K$ &$\chi^2$/DOF & $ F_{\rm 0.3-2\,keV}$ & $ F_{\rm 2-10\,keV}$ & $ F_{\rm 0.3-10\,keV}$    \\
(1) & (2) & (3) & (4) & (5) & (6) & (7)  \\
  \hline
35016002&	2.08(0.04)&	10.18(0.29)	&0.99/69&	3.16(0.10)&	2.34(0.14)&	5.50(0.15)\\
30376002&	2.27(0.04)&	4.47(0.13)&	0.92/51	&1.47(0.05)&	0.78(0.06)	&2.11(0.06)\\
30376003&	1.98(0.05)&	6.81(0.25)&	1.14/36&	2.06(0.09)&	1.79(0.16)&	3.85(0.16)\\
30376004&	2.02(0.05)&	7.66(0.29)&	1.04/34	&2.34(0.10)	&1.90(0.16)	&4.25(0.18)\\
 \hline \end{tabular} \end{minipage} \end{table*}

Furthermore, the parameter $E_{\rm p}$  showed a positive correlation also with the 2--10\,keV flux, which also was expected within the shock-in-jet scenario by (i) injection of high-energy electron population during the shock acceleration process (dominating in the observed X-ray variability over a large range of activity; \citealt{ac21}); (ii) dominance of synchrotron cooling of the highest-energy electrons over the IC "counterpart" \citep{z10}. Consequently, the source mainly followed  a trend of shifting the synchrotron SED peak towards higher energies with rising X-ray flux and conversely. On the other hand,  the mean cooling timescale of the X-ray-emitting electron population shortens  as $E_{\rm p}$ increases (by the both synchrotron  and IC mechanisms) and can compete with the acceleration timescales (see, e.g., \citealt{t09}). In turn, the latter effect may weaken of the  anti-correlation $b$--$E_{\rm p}$:  the cooling timescale becomes shorter than those of the stochastic and EDAP accelerations (yielding this anti-correlation). Note that these simulation results are in accord with a weak negative $b$--$E_{\rm p}$ trend for the subset with the positive $a-b$ correlation.

\subsubsection{Spectral trends and hysteresis patterns }

As discussed in Section\,4.2, the strong spectral variability of the source during 2005--2024  was mostly characterized by a HWB trend (generally observed for the HBL sources; see, e.g., \citealt{ah16,k22b,k23b}). This behaviour is interpreted in the framework of the shock-in-jet model, by taking into account the particle escape processes and synchrotron losses (\citealt{k98,s04} and references therein). Namely, during the shock propagation through the relativistic jet plasma, the density of the latter is locally enhanced and, consequently,  (i) the number of accelerating particles is boosted; (ii) increase in the magnetic field strength and establishment of the  higher particle acceleration rates expected; (iii) there is a shift of the maximum particle energies to higher values. As a result, a slope in the photon index--flux plane during X-ray flares is controlled by synchrotron cooling that, in turn, can explain (1) the observed anti-correlations of photon index--flux (shown by the both samples of the logparabolic and power-law spectra); (2) a stronger and faster variability in the number of the electrons producing the hard X-ray photons via the synchrotron mechanism. Namely, the $F_{\rm var}$ amplitude in the hard 2--10\,keV band is expected to be higher than that in the soft 0.3--2\,keV band, as exhibited by the source during the entire period 2005--2024 (61.8$\pm$0.6 and 49.6$\pm$0.3 per cents, respectively) and separate Periods\,1--7. However, the HWB trend was characterized by lower slopes  in some time intervals of diverse lengths (especially, in the highest states), or  the spectral evolution showed an unclear character and even the opposite trend on some occasions (as discussed above).  The latter was plausibly due to the addition of a new, soft flaring component to the emission zone. Consequently, the total XRT-band flux increased significantly at the expense of the soft 0.3--2\,keV emission and produced the highest-amplitude flares. The corresponding  data points produce outliers from the scatter plot photon index--flux and weakens the anti-correlation. 

Generally, a bluer-when-brighter (BWB) chromatism represents a lower-frequency extension of the HWB  trend  in the optical--UV energy range \citep{d23,k23b}. This chromatism was not statistically significant for our target in the case of the \emph{UVW2--U} bands, e.g.,  an  anti-correlation between the \emph{UVW2}-band magnitude and color-index \emph{UVW2$-$U} was detected below the adopted threshold of 99\,per cent, and only the index \emph{UVW2$-$B} showed an anti-correlation with the magnitude $m_{\rm UVW2}$ above this threshold (see Fig.\,\ref{figcor}k). The source showed a even stronger anti-correlation between $m_{\rm UVW2}$ and index \emph{UVW2$-$V}, as well as the BWB chromatism was detected in the case of the adjacent bands \emph{B} and \emph{V} (Figs\,\ref{figcor}l--\ref{figcor}m and Table\,\ref{cortable}). Similar to the HWB trend, some sub-samples  show a deflection from the BWB chromatism  (especially, during the highest optical--UV states), plausibly due to the emergence of a new, soft flaring component in the emission zone.

We also examined the spectral hysteresis patterns in the hardness ratio (HR)--flux  plane, which can provide us with some further clues about the acceleration mechanisms. The quantity HR was calculated as HR=$F_{\rm 2-10 keV}$/$F_{\rm 0.3-2 keV}$, where the unabsorbed fluxes were derived for the entire XRT observation in the case the latter was split into two or more segments. In this plane, the source may trace a clockwise (CW) spectral "loop", if the spectral evolution is governed by the flaring component starting in the hard X-rays. In turn, this component can be triggered by a rapid injection of ultra-relativistic electrons rather than by their gradual acceleration \citep{t09}. Such a situation can be implemented within the first-order Fermi mechanism,  by the Bohm's limit of particle diffusion (see, e.g., \citealt{tam09}). Fig.\,\ref{hyster}A presents three examples of the CW-type spectral loop, and the entire list of such instances shown by the source during Period\,1--7 is presented in Table\,\ref{hystertable}. Note that Fig.\,\ref{hyster}A3 shows two different, subsequent CW-loops completed  during a single X-ray flare occurring during MJD(570)40--62. Consequently, each loop could be related to the shorter-term flaring instances, which were not individually resolvable.

   \begin{figure*}  \vspace{-0.2cm}
 \includegraphics[trim=6.2cm 4.0cm 0.2cm 0cm, clip=true, scale=0.9]{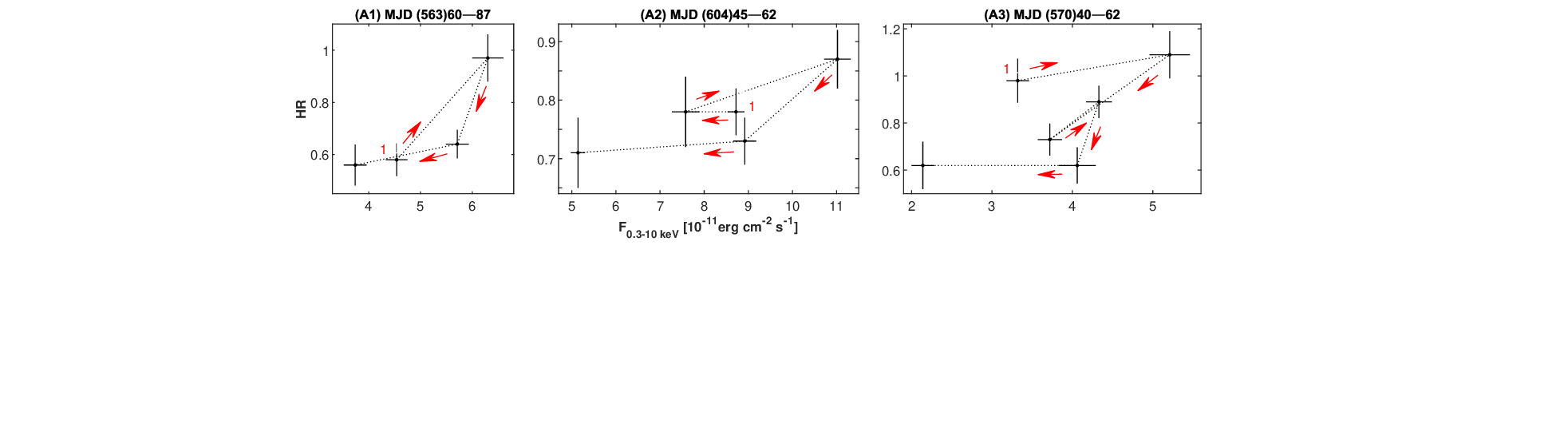}
 \includegraphics[trim=6.2cm 4.0cm 0.2cm 0cm, clip=true, scale=0.9]{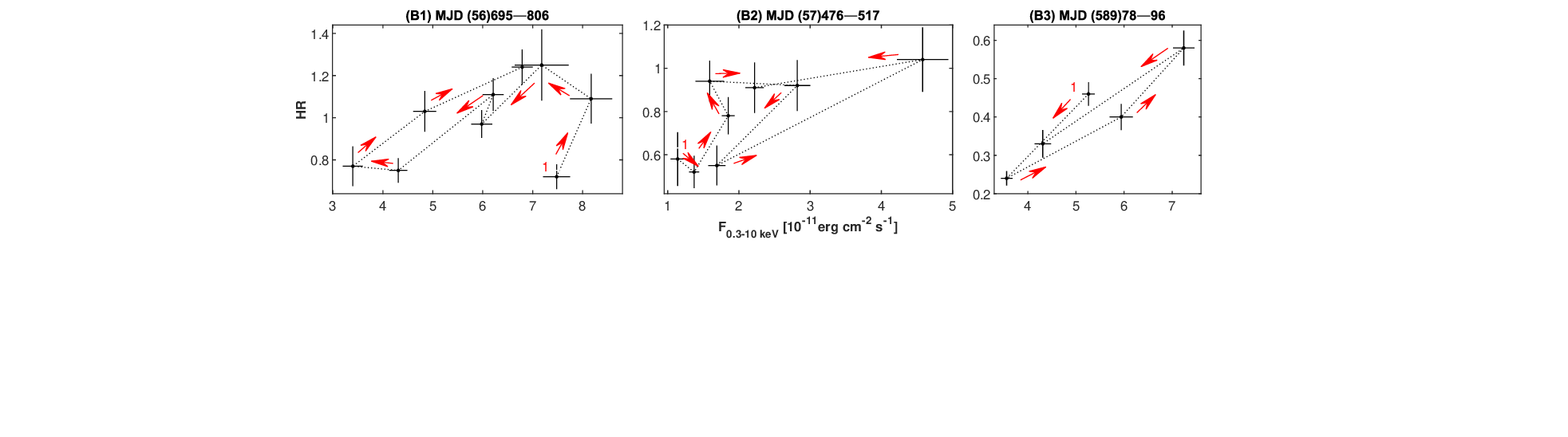}
 \includegraphics[trim=6.2cm 4.0cm 0.2cm 0cm, clip=true, scale=0.9]{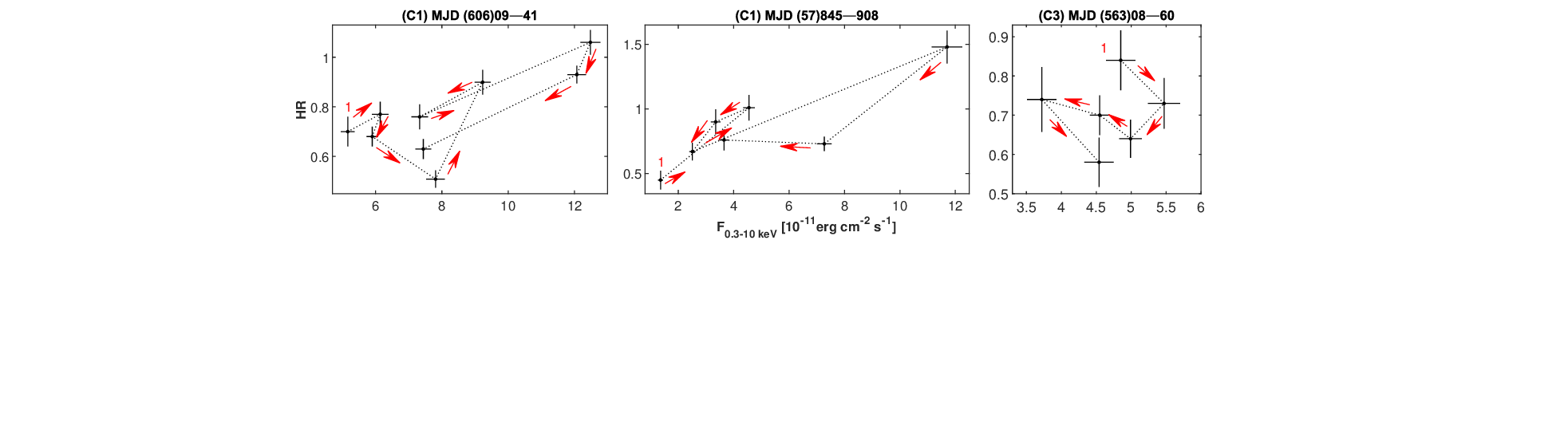}
\vspace{-0.7cm}
\caption{\label{hyster} Examples of the different hysteresis patterns during 0.3--10\,keV flares in the flux--HR plane: the CW and CCW-type loops are presented in rows (A) and (B), respectively; (C): transition from a CW-type loop into the opposite evolution and/or conversely.  In each case, the symbol "1" stands for the starting point and the subsequent "loop"  is traced by means of the red arrows.}
    \end{figure*}

The source frequently also showed  the counter-clockwise (CCW) loops (see Table\,\ref{hystertable} and Fig.\,\ref{hyster}B for the corresponding examples), to be observed when the specific X-ray flare propagates from low to high energies due to  gradual acceleration of electrons by the first-order Fermi mechanism (\citealt{f04} and references therein). In such a situation, the acceleration process cannot be instantaneous in the case of the significantly weaker (sub-Gauss) fields. Moreover, a slow, gradual acceleration and CCW-loops are also expected by the stochastic acceleration in the medium characterized by low magnetic fields and high matter densities \citep{vv05}. The first two plots of Fig.\,\ref{hyster}B show two subsequent CCW loops in each case. Several times, there was a CCW$\rightarrow$CW transition (or conversely) during the single X-ray flare. The most extreme case is presented in Fig.\,\ref{hyster}Ca: the source started with the CW-type spectral evolution, then changed into the opposite evolution twice during the 0.3--10\,keV flare that occurred during MJD(606)09--41 (the last flare in Period\,7; see also  Fig\,\ref{mwlper}Da). These results hint at the fastly variable and complex physical conditions in the X-ray emission zone of the target.

Moreover, the source did not show any clear hysteresis pattern  during some XRT-band flares (or their specific parts), possibly due to the co-existence of various acceleration mechanisms and complex physical conditions in the x-ray emission zone. Namely, electrons can undergo a complex acceleration scenario as follows \citep{k06}: initially, electrons are energized at the shock front via the Fermi-I mechanism, then injected into the shock downstream region and continue a further energization by the stochastic mechanism. After gaining sufficient additional energies, electrons can reach the  shock acceleration zone and repeat this cycle several times. Consequently, the source will not show any definite spectral loop if the number of such particles dominate among those producing X-ray photons.
Such a scenario could have another consequence: the curvature $b$$\sim$0.3 (efficient stochastic acceleration) may decrease in combination with the non-EDAP first-order Fermi process (yielding a power-law SED) and  weaken the anti-correlation $E_{\rm p}$--$b$ (as observed for 1ES\,1218$+$304).

\subsubsection{Signatures of the possible RMR}

Power-law photon spectra can be produced by the EED having the same functional shape:  $N(\gamma$)$\propto$$\gamma^{-p}$, with the so-called spectral-index $p$ related to the photon index $\Gamma$ as  $\Gamma$=$(p+1)/2$ (in the $\nu F_{\rm \nu}$ representation; see, e.g., \citealt{ah16}). Our spectral results yield the range $p$=2.2--4.4, with $p$$<$3 for the hard and very hard spectra. In turn, this required powerful acceleration processes in the relativistic jet of 1ES\,1218$+$304.  Moreover, the source  exhibited a much higher percentage of power-law spectra Compared to some nearby X-ray bright HBLs (more than 51\,per cent versus 10--28\,per cent shown by Mrk\,421  in different periods; see, e.g., \citealt{k20,k24b}). Only Mrk\,501 showed a comparable percentage of  the power-law spectra during some periods \citep{k23b}.   Consequently, the relatively frequent occurrence of power-law spectra in our target could be related to the higher importance of those acceleration processes, which yield a power-law EED. 

Note that the  significant portion of the target's power-law  spectra (at least 45 per cent, taking into account the associated errors) were hard or very hard ($\Gamma$$<$2 and $\Gamma$$<$1.80, respectively). In the simplest case, where particles are injected at the lowest possible energies, the competition between the acceleration and the escape forms a
power-law EED \citep{k06}. The simulations showed that the very and extremely hard power-law EEDs (yielding the correspondingly hard photon spectra) can be established when the acceleration is not dominated by escape and cooling (both synchrotron and IC) processes  (see \citealt{k06,st08}).

Different studies also demonstrated that a hard or very hard EED can be established by the relativistic magnetic reconnection (RMR)  in a turbulent, magnetized  jets medium (see, e.g., \citealt{sir13,sir14,b21}). According to our detailed spectral study, the target sometimes showed very fast logparabolic-to-powerlaw EED  transitions and conversely (e.g., between the spectra extracted from 330--700 second segments of a single XRT observation). Similar to the logparabolic-to-powerlaw and inverse EED transitions presented in Fig.\,\ref{recon}, the source showed comparable extreme instances during each of the three orbits of ObsID 30376104 (see Tables \ref{logp} and \ref{pow}). In order avoid a limitation of our analysis by considering only the XRT "window", the power-law spectra from these cases were re-fitted with the PLLP model by adding the simultaneous UVOT-band data, but no significant curvature was detected in these cases.

Note that source mostly showed a low spectral curvature ($b\sim$0.3 or lower) during logparabolic-to-powerlaw and converse transitions. As discussed above, such a curvature can be established by the turbulence-driven stochastic acceleration of the electrons producing the 0.3--10\,keV emission. As shown by  different simulations (see, e.g. \citealt{sir14,b21}), an RMR can be triggered by turbulence operating within a small-scale, magnetized jet area. Consequently, the target's X-ray emission zone sometimes could be the site of  turbulence-driven, fast reconnection in the small-scale, relativistically magnetized jet area  during those XRT observations. Namely, If we adopt the typical value of the jet bulk Lorentz-factor $\Gamma_{\rm em}$=10 (see, e.g., \citealt{f14}), a turbulence-driven RMR  could occur on the spacial scales of $d_{\rm tr}$= ${{ct_{\rm tr} \Gamma_{\rm em}}/({1+z}}$)=(1.6--3.6)$\times$10$^{13}$\,cm during the  most extreme cases. As for the transitions occurring during the different orbits, they could occur on larger spatial scales $\sim$10$^{14}$\,cm up to 10$^{15}$\,cm. These instances, plausibly caused by the RMR, should be characterized by various upstream magnetization to be lower than $\sigma_{\rm up}$ = 10, since the latter is expected for $p$$\lesssim$2 (i.e. $\Gamma$$\lesssim$1.5, never shown by the target). A number of similar instances have been reported also for Mrk\,421 and Mrk\,501, also explained by presence of turbulence-driven RMR \citep{k23b,k24b}.

The simulations showed that $p$$<$4  in the case of relativistic reconnection  versus $p$=4--11 in the non-relativistic regime, depending the strength of the so-called guide field (arising when the reconnecting current sheets do not exhibit perfectly anti-aligned magnetic fields; see \citealt{w21}). Since the latter corresponds to $\Gamma$$>$2.5, only one \textbf{possible} reconnection   instance could be performed in the non-relativistic regime (during  ObsID 303761).

Note also that the logparabolic-to-power-law EED  transitions and/or conversely occurred also when the source was showing curvatures with $b$$>$0.4. The latter is not expected in the case of efficient stochastic acceleration and, therefore, those transitions probably were lesser related to the turbulence-driven RMR but to subsequent shock passage through the jet area with the magnetic fields of  different confinement efficiencies (yielding the logparabolic and power-law EEDs, respectively). Consequently,  the magnetic field properties in the jet of 1ES\,1218$+$304  were significantly variable over the spatial scales of $\sim$10$^{13-14}$\,cm. The source sometimes showed a fast transition from the high to lower curvatures or vice versa over these spatial scales, indicating a fast change in the magnetic field properties (e.g., transition from the energy-dependent confinement efficiency into the turbulent one, or vice  versa).

As shown within some studies (e.g., \citealt{s18,h21}), a self-similar chain of plasmoids can be produced by a fast magnetic reconnection. The plasmoid interiors are compressed with time and amplify their internal magnetic field. In turn, this process will yield (i) energization of charged particles  by the magnetic moment conservation; (ii) the existing EED will extended to higher energies by a non-thermal tail $f(E)\propto E^{-3}$; (iii) the cutoff energy can increase with time as $E_{\rm cut}\propto \sqrt{t}$.  In the case of sufficient efficiency and duration of this process, it can supply the EED's highest-energy  tail with ultra-relativitic electron population capable of emitting in the BAT energy range and, in the most extreme cases, cause a shift of the synchrotron SED peak to the energies  beyond 10\,keV \citep{k24b}.   However, a connection-triggered plasmoid generation process was probably less important for 1ES\,1218$+$304 compared to some X-ray bright HBLs (e.g., Mrk\,421, Mrk\,501, 1ES\,0033$+$595):   our target never showed  $E_{\rm p}$$>$10\,keV and was detectable very rarely at the 5$\sigma$ confidence level with BAT.

\subsection{Flux variability }

\subsubsection{X-ray Variability on Various Timescales and the related physical processes}

During the long-term monitoring with the XRT, 1ES\,1218$+$304  showed a very different X-ray activity in diverse epochs: while the source underwent strong 0.3--10\,keV flares in 2012, 2018--2020 and 2022--2024, it mostly exhibited the low/medium XRT-band states and low-amplitude flares throughout the campaigns carried out during 2013--2017. Namely, the weakest activity was recorded in Periods 2 and 3, which were characterized by the lowest values of the parameters $F_{\rm var}$ and $F_{\rm mean}$, respectively (see Table\,\ref{xrtper}). At the moment of the highest historical 0.3--10\,keV  state (MJD\,59668, on 2022 March\,30),  our target was the 3rd brightest blazar in this energy range  (after Mrk\,421 and Mrk\,501) and by a factor $\sim$20 brighter than in the lowest recorded state. During the peaks of other strong XRT-band flares, 1ES\,1218$+$304 was also among the brightest extragalactic sources in X-ray sky.

  \begin{figure*} \vspace{-0.3cm}
\includegraphics[trim=6.1cm 1.8cm -1cm 0cm, clip=true, scale=0.9]{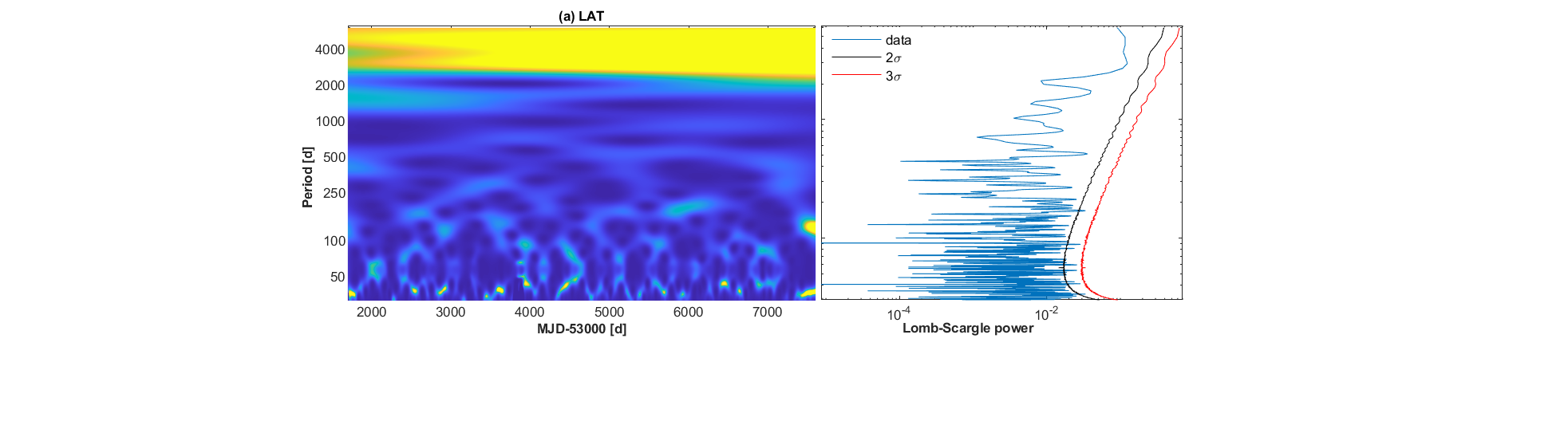}
\includegraphics[trim=6.1cm 2.0cm -1cm 0cm, clip=true, scale=0.9]{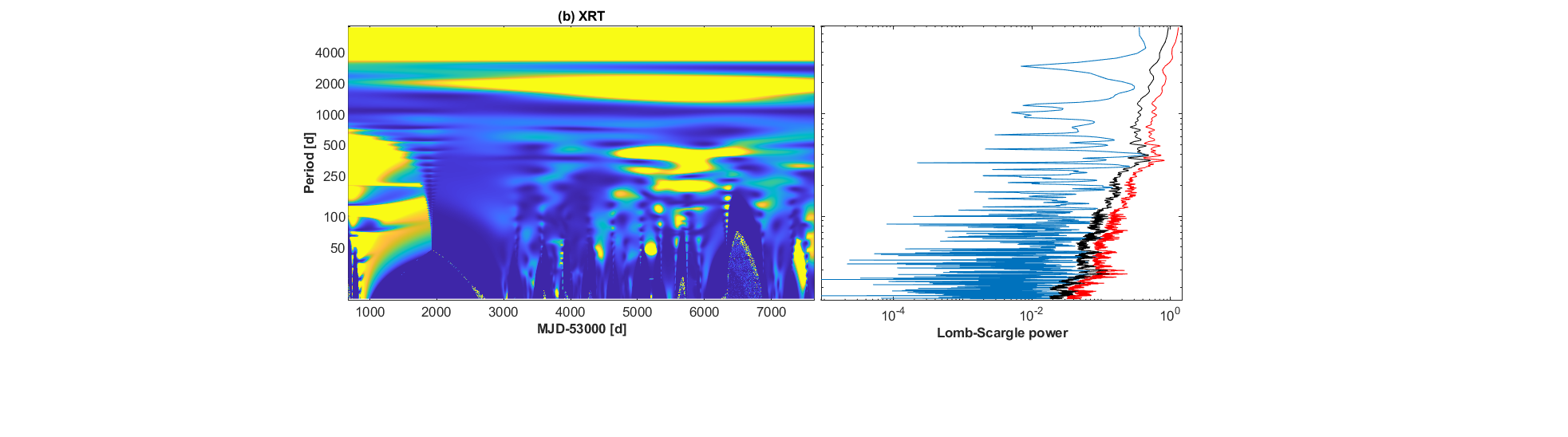}
 \vspace{-0.6cm}
 \caption{\label{lsp} The WWZ  and LSP plots:(a) four-weekly-binned 0.3--300\,GeV flux; (b) one-day-binned 0.3--10\,keV  flux.}
 \vspace{-0.4cm} \end{figure*}

We also checked the long-term MWL data sets for the possible presence of periodic flux  variability, which can be produced by the jet precession or periodical propagation of strong shocks through the jet (see, e.g., \citealt{tav08}). First of all, we performed this study for the most regularly sampled 0.3--300\,GeV light curve, constructed via the 4-week-binned flux values from the period 2008\,August--2023\,November, and  used the weighted wavelet Z-transform (WWZ; \citealt{f96} and references therein), which is one of the most commonly adopted technique for periodicity detection in blazars. The possible period is expected to emerge as a horizontal, narrow, permanent red strip in the WWZ plot (see, e.g., \citealt{o22}). In order to check the reliability of the possible periodicity and evaluate its significance, we also constructed a Lomb-Scargle periodogram (LSP; see \citealt{v18} for the latest review) for the same data train. In the case the aforementioned WWZ strip corresponds to the genuine period $P$, the corresponding peak should have a significance of   3$\sigma$ or higher in the LSP \citep{o22}. However, Figure\,\ref{lsp}a does not exhibit such detections, since all peaks are below the 3$\sigma$-curve. A similar situation is also present in the one-day-binned XRT and UVOT observations from the period 2005--2024 (Figure\,\ref{lsp}b). Note that  \cite{n18} did not found a periodicity from the long-term \emph{R}-band observations of 1ES\,1218$+$304. 

The duty cycle   of the 0.3--10\,keV IDVs  (defined as a fraction of total observation time during which the source the source was variable on intraday timescales;  \citealt{r99}) equals to 22.4\,per cent.  The latter is significantly lower compared to those  shown by nearby X-ray bright HBLs in this energy range during the different periods:  Mrk\,421 (43--83\,per cent; \citealt{k20,k24b}), Mrk\,501 (45--52\,per cent; see \citealt{k23b}), and 1ES\,1959$+$650 (40\%--52\%; \citealt{k18}). Note that this result could hint at the presence of weaker turbulence in the target's jet compared to these sources: in the framework of  the shock-in-jet scenario, the observed IDVs could be triggered by interactions of a propagating shock front with the jet inhomogeneities of turbulent origin  (see, e.g., \citealt{s04,m14,miz14}). However, a lower DC shown by 1ES\,1218$+$304 could be partially related to the fact that  these sources were generally much brighter (with significantly higher signal-to-noise ratios and, hence, easier to detected low-amplitude IDVs) than our target. Moreover, these HBLs were also characterized by much denser observational sampling on intraday timescales. Note that 1ES\,1959$+$650 showed  a comparable DC to that of our target in the periods 2005–2014 (28\%; \citealt{k16}) when the mean 0.3--10\,keV brightness and observational sampling was not significantly different from those of 1ES\,1218$+$304. A similar situation was also with 1ES\,0033$+$595, characterized by 26.5\,per cent during 2005--2022 \citep{k22b}. 

In the past studies, \cite{d23} reported a 0.3--7\,keV IDV  within 20\,min during the \emph{AstroSat} observations and LAT-band flux doubling with a timescale of 9.5\,hr. However, the latter was probably derived without an actual doubling of the 0.3--300\,GeV flux (by taking into account the uncertainties): our detailed study did not reveal such an instance, and the source was not even securely detectable with the time integrations shorter than 1\,d (see Section\,4.1). It's also evident that the source exhibited an unequal activity on intra-day timescales in different epochs and did not vary at the 3$\sigma$ confidence level during some extended multi-segment XRT observations with a total time span of 14--88\,ks (Fig.\,\ref{noidv}).

The source showed an anti-correlation between the fractional amplitude of the IDVs and and mean flux during the corresponding XRT observation (see Figure\,\ref{figcor}k and Table\,\ref{cortable}), explained by a strong non-stationary origin of the fastest X-ray variability (see, e.g., \citealt{z06}). Note that the majority of these instances occurred in the elevated X-ray states as demonstrated by Figure\,\ref{figcor}k. These results are in favour of the shock-in-jet scenario, explaining the IDV origin by the interaction of a propagating shock front with the jet inhomogeneities, particularly, with those characterized by higher matter density and stronger magnetic fields compared to the surrounding jet area \citep{m14}. As demonstrated by Table\,\ref{idvtable}, the source frequently showed a curved spectrum with $b$$\sim$0.3 or lower (expected in the case of efficient stochastic acceleration in the presence of a strong turbulence), or there was a transition from high spectral curvature to the lower $b$-value (a possible emergence of the small-scale turbulent structure in the X-ray emission zone) during the 0.3--10\,keV IDVs.  This result hints at the variable turbulence level with time, and relatively weak turbulence (leading to the lack/absence of small-scale inhomogeneities with strong magnetic fields) was probably the case in those epochs when the source was passive on intraday timescales.

During the 19\,yr period of the XRT monitoring,  the timescales of the fastest observed doubling/halving in the 0.3--10\,keV flux $\tau_{\rm d,h}$  ranged from 0.91\,d to 20.9\,d. Based on these  scales, it is possible to constrain the upper limit to the corresponding emission zone as $R_{\rm em}\leqslant {{c\tau_{\rm d,h} \Gamma_{\rm em}}/(1+z})$ \citep{sa13}. In the case of the most commonly-adopted value of the bulk Lorentz factor $\Gamma_{\rm em}$=10, we obtain a range of upper limits between   2.0$\times$10$^{16}$\,cm and 4.5$\times$10$^{17}$\,cm. However, the  XRT observations of 1ES\,1218$+$304 were not densely-sampled in those epochs when the source showed $\tau_{\rm d,h}$$>$10\,d and consequently, the corresponding upper limits could be significantly shorter. The fastest LAT-band flux doubling/halving (based on the robust detections of the target) were observed over timescales of 2--4 weeks.

\begin{table*}  \begin{minipage}{170mm}
  \caption{\label{klein} List of the time intervals characterized by spectral softening or hardening at the energies beyond 10\,GeV. Columns (3)--(5) present the number of the model-predicted photons, test-statistics and photon-index value in the 0.3--10\,GeV band, whereas the same quantities from the 10--300\,GeV band are provided in the Columns (6)--(8).}   \centering \vspace{-0.2cm}
   \begin{tabular}{cccccccccccc}        \hline
& &\multicolumn{3}{c}{0.3--10\,GeV} & \multicolumn{3}{c}{10--300\,GeV}  \\  
\hline
 Dates& MJDs  & $N_{\rm pred}$& TS  & $\Gamma$ &  $N_{\rm pred}$& TS  & $\Gamma$ \\
 (1)	& (2) &	(3)&	(4) &	(5)&	(6)& (7)& (8)\\
 \hline
 & & Softening&   \\ 
 \hline
2010 Jun 8—Sep 27&	(55)355--466&	78	&150	&1.43(0.16)&	8	&70	&2.16(0.18)\\ 
2011 Aug 30- Dec 19	&(55)803--914	&74&	142	&1.47(0.17)	&12&	108	&2.64(0.23)\\ 
2012 Jul 21--2013 Mar 1&	(56)129--352&	93&	88	&1.71(0.20)	&12	&113&	2.44(0.21)\\ 
2013 Mar 2---Oct 11	&(56)353--574	&73	&142	&1.20(0.20)&	8	&26	&1.75(0.20)\\ 
2016 Apr 5—Jul 25	&(57)483--594&	57	&56	&1.68(0.25)	&9	&69	&2.04(0.18)\\ 
2016 Nov 15--2017 Jun 26&	(57)707--930&	83	&149	&1.17(0.22)	&16	&122	&2.69(0.22)\\ 
2018 May 29—Sep 17	&(58)267--378&	62	&100	&1.46(0.20)&	9&	192	&1.67(0.14)\\ 
2019 Dec 10--2020 Feb 3	&(588)27--82&	67&	133	&1.43(0.18)&	10	&84	&2.21(0.20)\\ 
2020 Nov 10—2021  Jun 21&	(59)163--386&	214&	277&	1.78(0.12)&	14	&152&	2.20(0.15)\\ 
2022 Feb 1—Sep 11&	(59)611--832&	102&	169	&1.42(0.16)	&10	&91	&1.99(0.17)\\ 
2022 Sep 12—2023 Apr 23&	59833--60057&	98&	125&	1.59(0.16)&	10	&80&	2.05(0.18)\\ 
2023 Aug 15—Dec 4	&(60)171--282&	34&	52	&1.32(0.27)	&9	&91	&1.72(0.15)\\ 
\hline
& & Hardening &   \\
\hline
2008 Aug 5—Nov 24&	(54)683--794&	43&	22	&1.93(0.34)&	8	&67		&1.54(0.16)\\
2014 Sep 23—2015 Jan 12	&56923--57034&	75&	48	&2.21(0.36)	&9	&54		&1.73(0.16)\\
2017 Jun 27—Oct 16	&57931--58042&	105	&49	&2.46(0.40)&	10&	95		&1.79(0.15)\\
2017 Oct 17—2018 May 28	&(58)043--266&	210	&259	&1.79(0.12)&	15	&132		&1.47(0.13)\\
2020 Mar 31—Jul 20	&58939--59050&	65&	67	&1.94(0.26)	&9	&92	&	1.62(0.14)\\
  \hline \end{tabular} \end{minipage} \end{table*}

\subsubsection{Lognormal variability and MWL correlations}

We checked that the distribution of the unabsorbed 0.3--10\,keV flux derived from all XRT observations of 1ES\,1218$+$304 rather close to the lognormal distribution than to the Gaussian shape. Most commonly, this result is explained an imprint of  those instabilities onto the jet which occur in the blazar accretion disc and trigger the shock waves propagating through the jet (see, e.g., \citealt{r19,k24b}). However, the lognormal function does not yield a better fit with the sample of the flux values from those XRT observations of  the target  characterized by the 0.3--10\,keV IDVs. This result can be explained by  a higher fraction of 0.3--10\,keV  emission from the ultra-relativistic electron populations energized by the local, jet-inherent instabilities.

The distribution of the 0.3--300\,GeV flux  also showed a preference of the lognormal function over the Gaussian one. Note that Fig.\,\ref{figcor}o and Table\,\ref{cortable} show a  positive correlation between the 2-week-binned XRT and LAT-band fluxes. The intrinsic $F_{\rm 2-10 keV}$--$F_{\rm 3-300 GeV}$ correlation could be stronger since the source frequently showed the low X-ray and HE states simultaneously. However, the two-weekly integrated LAT data frequently did not yield the robust detections of the source in the lower 0.3--300\,GeV states, and the calculated upper limits were not used for the correlations study.  Moreover, the intrinsic correlation could be underestimated due to using the 2-week  integrations in the both bands: the source frequently showed a large variability during this time interval. However, we could not use narrower bins owing to the problem with a robust LAT-band detection of the target (less than 30\,per cent for one-week-binned data, biased to the flaring LAT-band states). 

The positive $F_{\rm LAT}-F_{\rm XRT}$ correlation could be caused by the IC-upscatter of the X-ray photons to the LAT energy range in the KN regime    (see, e.g., \citealt{k23b,k24b} presenting similar results for Mrk\,421 and Mrk\,501). As shown in Appendix\,A, short-term LAT-band flares were frequently seen in the epochs of the XRT-band "counterparts" hinting at the possible connection between these instances. Generally, the KN-regime is characterized by a suppression of the $\gamma$-ray output and, consequently, the $F_{\rm var}$ quantity was almost 50\,per cent lower in the entire LAT band than that in the 2--10\,keV band for the entire data sets (as well as in some periods). Moreover, the existence of a KN-suppression is in agreement with the target's general faintness in the  0.3--300\,GeV energy range. Note that Table\,\ref{klein} also presents  those time intervals when 1ES\,1218$+$304 showed a softening at the energies beyond 10\,GeV compared to the lower-energy LAT-band part of the spectrum. These instances can be explained by possible IC-upscatter of  X-ray photons to the GeV energies in the KN-regime (versus the upscatter of lower-energy photons in the Thomson regime to the energies below  10\,GeV). Consequently, the possible KN-suppression weakened the expected strong correlation between the fluxes extracted in the  0.3--10\,GeV and  10--300\,GeV band down to $\rho$=0.50$\pm$0.10  (Fig.\,\ref{figcor}p), as well as yielded a lower $F_{\rm var}$ value in the latter band compared to the lower-energy one (33.4$\pm$3.1 and 39.7$\pm$2.8 per cents, respectively).

According to \cite{t11}, one should observe a 
change in the SED curvature parameter at the energy corresponding to the transition from the the TH to the KN regime. Although the source was characterized by curved LAT-band spectra in the corresponding time intervals (see Section\,4.2.2), the parameter $b$ showed low values and it is not possible to detect the spectral curvature in the both separate 0.3--10\,GeV and 10--300\,GeV bands at the significance of 2$\sigma$ and higher owing to target's HE faintness. Nevertheless, the upscatter of optical--UV photons in the TH regime (dominating in the 0.3--10\,GeV band) could make some contribution although at  the energies beyond 10\,GeV. The latter possibility is in agreement with the general presence of the $\gamma$-ray SED peak at the energies higher than 100\,GeV for during the aforementioned time intervals and with the results of the broadband SED modelling, which showed 1ES\,1218$+$304 to be a TeV-peaked candidate source \citep{c18,c20}.  

The  possible upscatter of optical--UV photons to the MeV--GeV energies in the TH regime is reflected in the positive correlation between the LAT and UVOT-band emissions (see Fig.\,\ref{figcor}q and Table\,\ref{cortable}).  The data points corresponding to the time intervals provided in  Table\,\ref{klein} make outliers from the scatter plot $F_{\rm 3-300 GeV}--\Gamma$ (Fig.\,\ref{figcor}j), which show a dominance of the softer-when-brighter spectral evolution of the source during the LAT-band flares and hints at the importance of upscattering the optical--UV photons in the TH regime (see, e.g., \citealt{k25}).

Note that the  weakness of the correlation $F_{\rm 2-10 keV}$--$F_{\rm 3-300 GeV}$  could be also caused by a significant contribution from those 0.3--300\,GeV photons produced by the cascade-related emission processes (e.g., synchrotron radiation from the proton-induced hadronic cascades). The latter are also notable for  producing a lognormal variability \citep{r19} and very/extremely hard $\gamma$-ray spectra (see, e.g., \citealt{m93,sh15}). Such spectra were  frequently exhibited by the target in the 0.3--300\,GeV energy range (as demonstrated in Fig.\,\ref{figdistr}e). Generally, a correlation between  X-rays (synchrotron emission from ultra-relativistic electrons in HBLs) and  $\gamma$-ray emission from the proton-induced hadronic cascades is not expected,  a significant hadronic contribution to the total 0.3--300\,GeV energy "budget" could weaken  the  observed $F_{\rm XRT}-F_{\rm LAT}$ correlation. Note that Table\,\ref{klein} also presents  those time intervals when the source showed a hardening at the energies beyond 10\,GeV compared to the lower-energy $\gamma$-ray emission. However, the detections of the neutrino-related events (inherent to the hadronic cascades; see, e.g., \citealt{c15}) have not been reported for 1ES\,1218$+$304 to date and, therefore, we can not draw firm conclusions about the importance of the hadronic processes in the jet our target.

Although the optical--UV fluxes showed a better fit with the logparabolic function than with the Gaussian one, this difference progressively decreased towards the longer wavelengths, possibly owing to the increasing contribution of the emission produced in the different jet regions and related to the local unstable processes. Consequently, the \emph{UVW2}-band variability showed a gradually weaker correlation with those shown in other optical--UV bands, in the direction \emph{UVM2}$\rightarrow$\emph{R}  (Figs.\, \ref{figcor}q--\ref{figdistr}r and Table\,\ref{cortable}). Especially notable were two strongest optical--UV flares occurring during 2018--2020 (Period\,5), while the simultaneous 0.3--10\,keV flares were  not the strongest during 2005--2024. We suggest that there were additional electron populations during the both flares, which were energized by some local unstable processes and producing photons predominantly in the optical--UV energy range.  Consequently, the highest historical states were observed in the UVOT and \emph{R} bands (in contrast to  X-rays) and produced  outliers from the from the scatter plot showing a positive correlation  $F_{\rm 0.3-2 keV}$--$F_{\rm UVW2}$ (Fig.\, \ref{figcor}s).  Non-correlated optical--UV and X-ray variabilities were observed also some parts of Periods\,1--3 and demonstrated the requirement of  multi-zone scenarios.

\section{Summary and Conclusions}
In the framework of this study, we present the results of long-term MWL observations carried out for the X-ray-selected BL Lac source 1ES\,1218$+$304 during 2005 October–2024 November. The MWL data set was obtained with the $\emph{Swift}$-based X-ray and optical--UV instruments, as well as with different space missions and ground-based telescopes. Our basic results and the corresponding interpretations:
\begin{itemize}
\item In the 0.3--10\,keV  band, the source exhibited strong flux and spectral variability, with the highest and lowest states differing by a factor of $\sim$20 in the brightness. The strongest XRT-band flares were recorded in the period 2018--2023, when the unabsorbed flux was sometimes higher than 2$\times$10$^{-10}$\,erg\,cm$^{-2}$\,cm$^{-1}$ and 1ES\,1218$+$304 became one of the brightest blazars in the X-ray sky during those observations. At the moment of the highest historical 0.3--10\,keV  state,  our target was the 3rd brightest blazar in this energy range. On the contrary, it mostly exhibited the low/medium XRT-band states and low-amplitude flares the campaigns carried out during 2013--2017. On average, a similar behaviour was observed in the optical--UV and  0.3--300\,GeV energy ranges, although the highest UVOT and \emph{R}-band states were recorded  during two strongest flares in those epochs when the source did not show the highest 0.3-10\,keV brightness (possibly due to the emergence of additional electron population in the emission zone, energized by some local unstable processes and emitting predominantly at the optical--UV frequencies). The source was faint and rarely detectable at the 5$\sigma$ confidence level with MAXI and BAT. Although showing a strong flaring activity on diverse timescales, 1ES\,1218$+$304 generally was not a bright source in the LAT band and never was detected above the level of 10$^{-7}$ph\,cm$^{-2}$s$^{-1}$) in the  0.3--300\,GeV energy ranges. Along with the distance, the LAT-band faintness can be explained by possible  IC-upscatter of X-ray emission at the energies beyond $E$$\gtrsim$10\,GeV in the KN-regime, leading to the suppression of the $\Gamma$-ray outcome and frequent observation  of the correlated XRT and LAT band flaring activities. On long-term timescales, the source showed a non-periodic and lognormal variability  XRT, UVOT, LAT, \emph{R}-band fluxes, which hints at imprinting of the disc unstable processes onto the jet.

\item During the relatively intense XRT campaigns, the source showed 12 instances of flux doubling/halving with timescales ranging from 0.91\,d to more than 10\,d and constraining the upper limit to the variable emission zone as  2.0$\times$10$^{16}$\,cm--4.5$\times$10$^{17}$\,cm. Moreover, 1ES\,1218$+$304 varied 38-times on intraday timescales, with fractional variability amplitudes of 3.6(0.8)--31.5(6.6) per cent and most frequent occurrence during the elevated and flaring 0.3--10\,keV  states (compatible with the shock-in-jet scenario). While the majority of these IDVs occurred on subhour timescales (and 12 times within 1-ks exposures down to 540\,s), some densely-sampled XRT campaigns  were characterized by only slow variability of the target,  without IDVs and did not show the flux doubling/halving instances.  This result hints at the variable turbulence strength, leading to the lack or even absence of small-scale inhomogeneities with strong magnetic fields (required to produce extreme brightness changes via the interaction with the moving shock front) in some epochs.

\item More than 51\,per cent of the 0.3--10\,keV spectra did not show curvature and were well-fit with a simple power-law, and this result is not common for the HBL sources. Frequently, these spectra were (very) hard, and the source showed very fast   logparabolic-to-powerlaw EED  transitions and/or conversely for the spectra extracted from the 330--700 second segments of a single XRT exposure or from those corresponding to the different segments of the multi-orbit \emph{Swift} observations. These features are most naturally explained by means  of the possible turbulence-driven RMR occurring in jet area with a small spatial extent of $\sim$10$^{13-14}$ cm. In contrast to some HBLs, 1ES\,1218$+$304 showed spectra with relatively large curvature ($b$$>$0.4), very hard photon-index at 1\,keV ($a$$\sim$1.8--1.5) and an anti-correlation $a$--$b$ in some epochs, to be established by the EDAP-acceleration of the electron population with a very low initial energy distribution. Other spectra with the relatively large curvature were characterized by the positive  $a$--$b$ correlation and negative $b$--$E_{\rm p}$ trend, expected withing the EDAP scenario when the initial EED is not very low. The latter correlation was shown also by the another sample of curved spectra were characterized also by curvatures of $b$$\sim$0.3 or lower, expected in the case of  efficient stochastic acceleration. Sometimes, the source  showed a fast transition from high to low curvature (or conversely), indicating a fast change in the magnetic field properties over small spatial scales: transition from the magnetic field with the energy-dependent confinement efficiency into the turbulent one, or vice versa.

\item Due to the possible presence of the different acceleration mechanisms, the source showed the  the relation $S_{\rm p}$$ \varpropto$$E^{\alpha}_{\rm p}$ with the value of the exponent relatively close to $\alpha$=1.5, when the mean energy and number of the X-ray emitting electrons are variable. Moreover, the source mainly followed a  HWB spectral evolution  (similar to other HBLs), expected by the injection of hard EED during the 0.3--10\,keV flares and dominance of synchrotron cooling of the highest-energy electrons over the IC-cooling. However, the source showed a deflection from this general trend during some time intervals, possibly, due to the emergence of a new, soft flaring component in the X-ray emission zone. These instances and weakness of the UV--X-ray--$\gamma$-ray correlations demonstrate that the one-zone SSC model was not always valid for the target, requiring more complex emission scenarios. 
\end{itemize}

\section{Data availability}
For this work, we have used data from $\emph{Swift}$-XRT, $\emph{Swift}$-UVOT, $\emph{Swift}$-BAT, MAXI, \emph{Fermi}-LAT, \emph{RXTE}-ASM, Tuorla Blazar Monitoring Program, KAIT Fermi AGN Light-Curve Reservoir, Steward Observatory and  Catalina Real-time Transient Survey, which are available in the public domain. The corresponding details are provided in Section\,3.

\section*{ACKNOWLEDGEMENTS}
BK and AG thank Shota Rustaveli National Science Foundation and E. Kharadze National Astrophysical Observatory (Abastumani, Georgia) for the fundamental research grant FR$-$21--307 and Andrea Tramacere for his useful suggestions.  We acknowledge the use of public data from the \emph{Swift} data archive. This research has made use of the \texttt{XRTDAS} software, developed under the responsibility of the ASDC, Italy.  In our study, the data from the Steward Observatory spectropolarimetric monitoring project
supported by Fermi Guest Investigator grants NNX08AW56G, NNX09AU10G,
NNX12AO93G, and NNX15AU81G) were used.   The Catalina survey is supported by the U.S.~National Science Foundation under grants AST-0909182 and AST-1313422. We thank the anonymous referee for his/her useful suggestions.

\label{lastpage}

\begin{appendix} 
\section{MWL variability in different periods and Details of Subhour 0.3--10\,keV IDVs}

Below, we summarize the most important results from the X-ray and MWL timing study from Periods\,1--7, the time ranges and summary of which are provided in Tables \ref{xrtper} and \ref{uvotper}. 
\vspace{-0.2cm}
\begin{itemize}
\item  The highest historical 0.3--10\,keV  state, corresponding to about 2.7$\times$10$^{-10}$\,erg\,cm$^{-2}$\,cm$^{-1}$, was recorded in  Period\,6 (see Fig\,\ref{mwlper}A). Unfortunately, 1ES\,1218$+$304  was observed only once with \emph{Swift} in that epoch, and the subsequent XRT visit to the  the source occurred after almost 4\,months later when it was showing a flaring state again: the XRT-band flux was about twice higher than that from all XRT observations of the target. The second half of the period was characterized by the most densely-sampled \emph{Swift} campaign performed during 2005--2024 (43 visits to the source  in 185\,d) and by long-term X-ray flare by a factor of $\sim$4.5 lasting longer than 5\,months: the flare was not ceased when the monitoring was interrupted owing to the target's seasonal invisibility. While the source varied relatively slowly on timescales of a few days, there was a flux doubling instance  at the flare onset with the timescale $\tau_{\rm d}$=13.2\,d.  This instance was preceded by a flux halving with the timescale $\tau_{\rm d}$=7.6\,d. The peak brightness was achieved via the short-term flare, incorporating a brightness increase by about 80\,per cent within 8\,d. During the entire period, the source was not detected at the 5$\sigma$ confidence level with BAT and MAXI. Although a strong flaring state was recorded in each of the UVOT bands  along with the highest XRT-band flux, this was not the highest level shown by the source at the optical--UV frequencies during 2005--2024 (about 50--60 per cent lower; see below). Moreover, a long-term flare was recorded also in these bands during the period's second half (along with the 0.3--10\,keV "counterpart", with the 4--5 day delay in the peak brightness), but the amplitudes were significantly lower and showed a declining trend towards the longer wavelengths. A weak flaring activity was recorded with the LAT, characterized by elevated 0.3--300\,GeV states along with the highest 0.3--10\,keV fluxes in the periods first and second halves, respectively.

\item The source was detected flaring during the XRT campaign performed around MJD\,59490 (Period\,5): the 0.3--10\,\,keV flux showed two subsequent fast fluctuations by $\sim$75\,per cent around the level of 4.5\,cts\,s$^{-1}$ (see Fig\,\ref{mwlper}B). The latter was about twice higher than the mean rate during 2005--2024. A similar behaviour and comparable fluxes were recorded also in the periods second half (around  MJD\,59875). Between these instances and at the period's end, the source showed the states with CR$\sim$2\,cts\,s$^{-1}$ and lower. During the second flare, the source showed one of the strongest LAT-band flare (by a factor of $\sim$6), while there was a lower-amplitude flare during the first, relatively strong 0.3--10\,keV instance, and the source simultaneously showed lower X-ray and $\gamma$-ray states. On the contrary, the strongest optical--UV flare and  highest historical states in the UVOT  and \emph{R}-band states were observed at the period's end. Almost comparable states and another strong optical--UV flare occurred along with the first strong X-ray instance. Again, BAT and MAXI did not detect the source with 5$\sigma$ significance.

\item Another strong 0.3--10\,keV flare occurred  around MJD\,56064  (Period\,1) with $\sim$4.4\,cts\,s$^{-1}$ (corresponding to the unabsorbed flux of about 1.5$\times$10$^{-10}$erg\,cm$^{-2}$s$^{-1}$; see Fig\,\ref{mwlper}C), recorded in the course of the flux-doubling instance with $\tau_{\rm d}$=4.2\,d. In this periods, the source was characterized by 5$\sigma$-detections with BAT and MAXI, as well as by strong 0.3--300\,GeV flare (by a factor of $\sim$5) along with the strongest MAXI-band activity. In that epoch, the source showed flares also in the UVOT  and \emph{R}-bands.

\item Two strong flares were recorded in the beginning and end of Period\,7, which were not observed entirely and could have even higher amplitudes (Fig\,\ref{mwlper}D). Simultaneously, the source showed elevated 0.3--300\,GeV and optical--UV flares. Unfortunately, no \emph{Swift} monitoring was carried out around MJD\,60550, when the source showed an enhanced MWL activity in LAT, BAT and MAXI bands.

\item Two strong XRT-band flares by factors of 3--5 occurred in the first half of Period\,4 (Fig\,\ref{mwlper}E). The peak of the second one was preceded by flux doubling and halving instances with $\tau_{\rm d,h}$=8.8--13.9\,d. A flux-halving instance with $\tau_{\rm d,h}$=20.9\,d was recorded also after the peak of the first flare. The third flare occurred at the period's end, but only four XRT observations of 1ES\,1218$+$304 were carried out and it's impossible to draw a firm conclusion about the amplitude. The source was characterized by an elevated MAXI-band activity and 5$\sigma$-detections. While the low-amplitude LAT-band flares were recorded during each 0.3-10\,keV "counterparts", the strongest $\gamma$-ray instance by a factor of $>$5.5 occurred around MJD\,58008 when the source was not monitored with \emph{Swift}. Each XRT-band flare were accompanied by those at the optical--UV frequencies. 

\item The source underwent a strong 0.3-10\,keV flare at the end of Period\,2, and a similar behaviour was observed also with LAT and UVOT (Fig\,\ref{mwlper}F). The 0.3-300\,GeV flare was preceded by another instance with the same amplitude when the source was not targeted by \emph{Swift}. The period's start was characterized by lower-amplitude, fast XRT-band flares by 40--60\,per cent, accompanied by elevated  optical--UV and LAT-band states. During the both epochs, the source showed an elevated  MAXI-band activity and 5$\sigma$-detections.

\item Finally, the lowest 0.3-10\,keV states and  weakest flaring activity were recorded in Period\,3, when the strongest flare (observed during MJD\,57434--57476) incorporated flux doubling and halving instances with $\tau_{\rm d,h}$=3.1--11.7\,d. This flare was preceded and followed by the flux-halving ($\tau_{\rm h}$=10.2\,d) and doubling ($\tau_{\rm d}$=3.8\,d) instances. The fastest flux-halving occurred in the period's first half (on the intraday timescale, MJD(5707)5--6). In the epochs of the XRT observations, the source showed low-amplitude flares  in the optical--UV and 0.3-300\,GeV energy ranges.
\end{itemize}

The details of the subhour 0.3-10\,keV  IDVs were as follows: the fastest  instance showed the subsequent flux drop and increase  with $f_{\rm var}$=12.6(2.47) during the 0.5-ks segment of ObsID\,30376162 (MJD\,60128; the second orbit). The opposite cadence was shown during the first orbit of the same observation, but the variability was below the aforementioned threshold and, therefore, not listed in Table\,\ref{idvtable}. The subsequent brightness drop and increase within the 1-ks exposure occurred also during the IDVs presented in Figs\,\ref{idv}b--\ref{idv}h. A similar behaviour was evident during the longer IDVs presented in Figs \ref{idv}l, \ref{idv}p and \ref{idv}s--\ref{idv}t. These instances were characterized by $f_{\rm var}$=8.4(2.2)--31.5(6.6) per cent.  

Note that one of these instances (detected during the second orbit of ObsID\,30376127; Fig.\,\ref{idv}) was preceded by another IDV when the 0.3-10\,keV flux increased at least 50\,per cent within 0.72\,ks. On the contrary, the brightness almost halved within 0.96\,ks in the case of ObsID\,30376069 (MJD\,57487; Fig.\,\ref{idv}j). The IDVs with the subsequent brightness increase and decline are presented in Figs\,\ref{idv}i--\ref{idv}k and \ref{idv}r. The most extreme behaviour was observed during the first orbit of ObsID\,30376098 (lasting 1.2\,ks; Fig.\,\ref{idv}m): after the initial decline, the source showed a subsequent brightening and decline cycle with an almost symmetric flare profile and the subsequent brightening. However, the source was not variable with  3$\sigma$ significance during the second orbit of the same observation, performed after about 3.5\,ks.
\clearpage

\onecolumn
\section{XRT observation with most extremely fast transitions between the logparabolic and power-law spectra}
 \begin{figure*}
\includegraphics[trim=6.2cm 1.0cm 0cm 0cm, clip=true, scale=0.91]{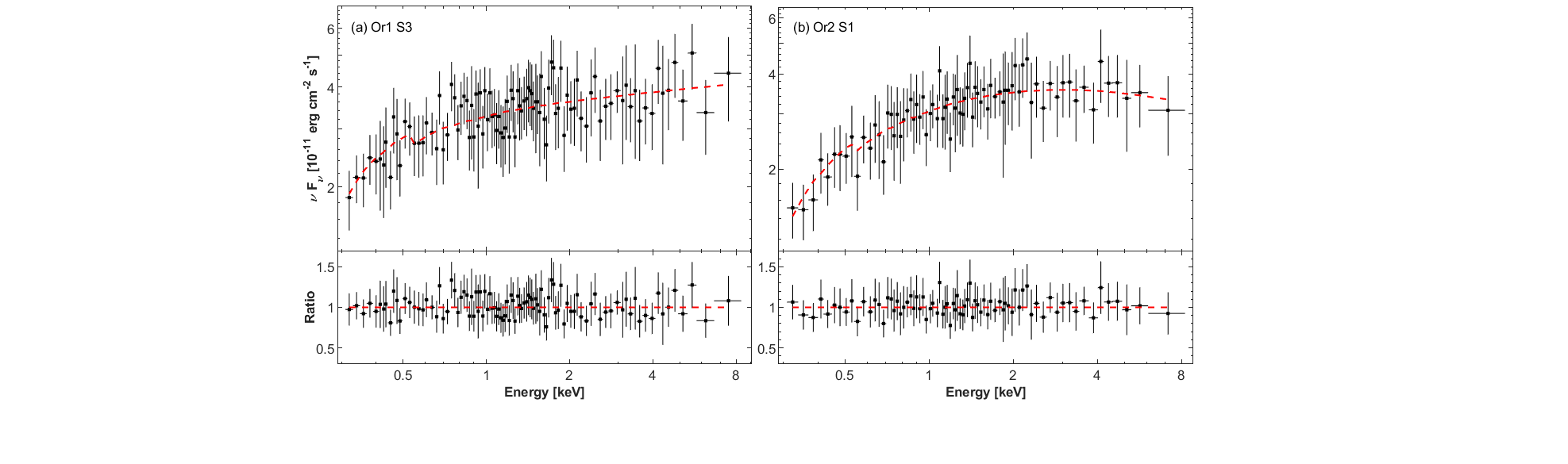}
\includegraphics[trim=6.2cm 1.3cm 0cm 0cm, clip=true, scale=0.91]{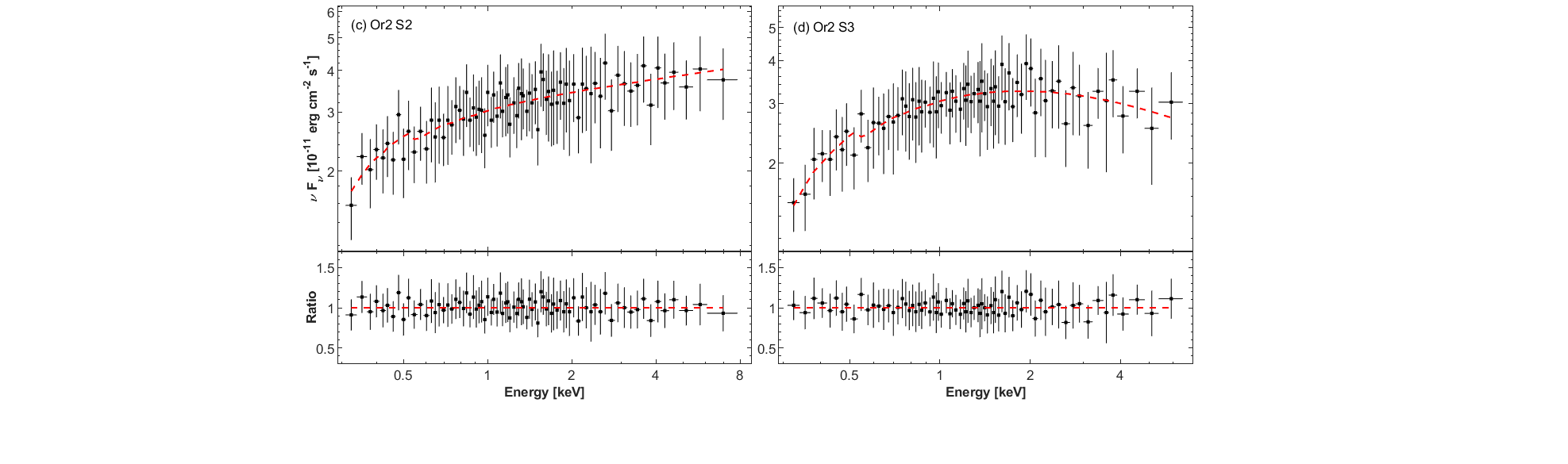}
\vspace{-0.7cm}
 \caption{\label{recon} The 0.3--10\,keV spectra extracted from the four different segments of ObsID\,30376015, showing a consecutive change from the power-law into the logparabolic SED and conversely: the 3rd 735-second segment of the first orbit and three subsequent 600-second segments of the second orbit. While the dashed red curve represents the power-law model in panels (a) and (c), it stands for the  logparabolic model in panels (b) and (d). For each spectrum, the fit residuals are presented in the bottom panel.}
 \end{figure*}

 \clearpage
\section{Examples of XRT observations without IDVs and full list of hysteresis patterns}
\begin{figure*}
\includegraphics[trim=6.0cm 5.3cm 0cm 0cm, clip=true, scale=0.91]{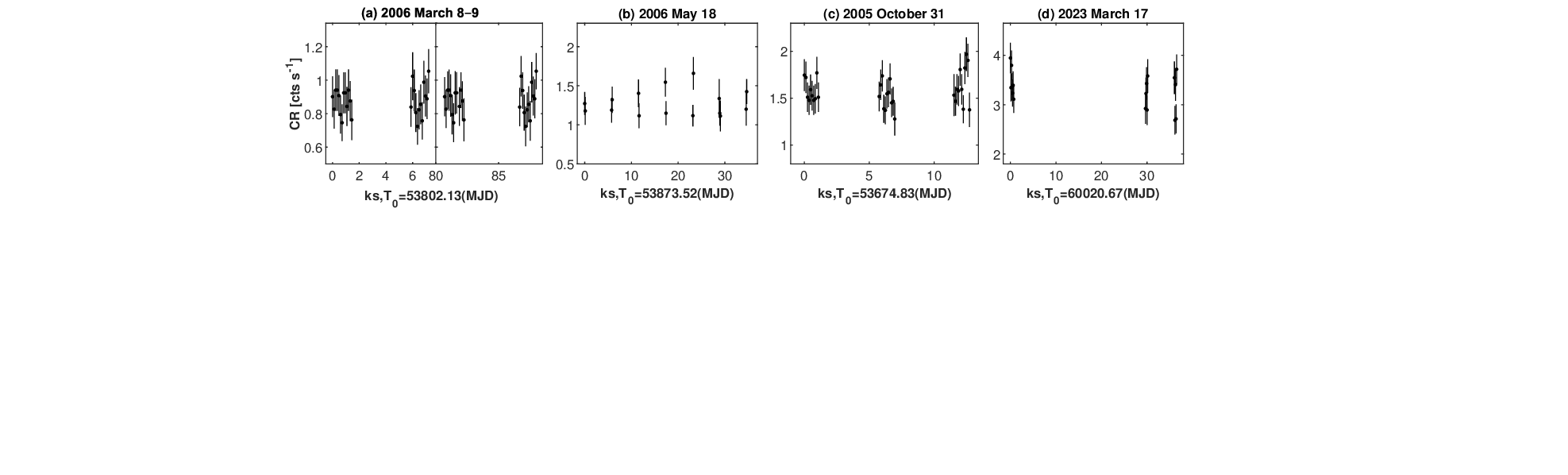}
\vspace{-0.9cm}
 \caption{\label{noidv} The relatively densely-sampled XRT observations of 1ES\,1218$+$304 not showing the 0.3--10\,keV IDVs.}
 \end{figure*}

 \begin{table*}  \vspace{-0.3cm}  \small \centering
   \begin{minipage}{180mm}
  \caption{\label{hystertable} List of the CW and CCW-type loops in the flux--HR  plane (see Fig\,\ref{mwlper} for the instances described in the last column).}
     \vspace{-0.2cm} \centering
  \begin{tabular}{cccc}
  \hline
 Period & MJDs & The XRT-band Variability Instance \\
 \hline 
    &  & \multicolumn{1}{l}{CW-loops} \\
      \hline
1  & (560)00--41 & Period's strongest flare   \\  
2  & (563)08--30 & First short-term flare \\
2& (563)60--87  &  The third, relatively strong flare\\
3& (570)40--76 & Strongest flare in the period's first half\\
3& (570)40--76 & Last low-amplitude flare in the period's first half\\
3& (57)349--430 & First flare in the period's second half\\
3& (574)51--459 & Period's strongest flare\\
4  & (58)870--908 & Period's strongest flare\\
5& (584)82--92 & First short-term flare \\
6 & 59992--60013  & Short-term, low-amplitude flare at the onset of the long-term one \\
6 & (600)13--44  & Brightening phase of the long-term flare\\
6 & (600)62--68  & Short-term flare in the maximum epoch of the long-term flare\\ 
7 & (604)45--62  & First flare\\ 
7 & (606)29--41  & Second flare\\
     \hline
 &   & \multicolumn{1}{l}{CCW-loops}\\
     \hline
1  &55973--56000  &  First long-term, low-amplitude flare \\   
2  & (563)30--60 & The second, low-amplitude flare\\
2& (56)390--436 & Last three short-term, low-amplitude flares in the period's first half \\
2 & (56)695--769 & Period's strongest flare \\
3  & (57)076--107 & Short-term, low-amplitude after the strongest one \\
3& (574)59--70 & Declining phase of the strongest flare\\
3& (574)59--70 & Low-amplitude flare after the strongest one\\
3& (57)491--517 & Period's last flare\\
4& (577)56--83 & Period's first flare\\
4 & (57)783--845 & Declining phase of the period's first flare\\
4 & (578)45--70 & Short-term, low-amplitude flare superimposed on the longer-term one \\
4& (59)695--706 &Period's  last  flare\\
5& (58)492--500 & Second short-term flare \\
5& (588)52--54 & Strongest flare in the period's second half\\
5& (58)492--500 & Low-amplitude flare at the period's end \\
6 & (600)44--55  & Low-amplitude brightness fluctuation in the maximum epoch of the long-term flare\\
6 & (60)068--156  & Short-term flare and subsequent brightness fluctuations in the maximum epoch of the long-term flare\\ 
7 & (606)13--29  &Period's second flare\\ 
   \hline  \end{tabular} \end{minipage} \end{table*}
\end{appendix}

\end{document}